\documentclass[english,aps,prl,amsmath,twocolumn,preprintnumber,superscriptaddress,notitlepage]{revtex4-2}
\usepackage[T1]{fontenc}
\usepackage[latin9]{inputenc}
\makeatletter

\usepackage{mathptmx,newtxtext,newtxmath,xspace}
\usepackage{amsbsy,bm,bbold}
\usepackage{graphicx,color,xcolor,epsfig,rotate}
\usepackage{fancyhdr}
\usepackage{soul}

\makeatother

\usepackage{babel}
\begin{document}
\title{Renormalized Classical Theory of Quantum Magnets}
\author{David A. Dahlbom}
\email{ddahlbom@utk.edu}
\affiliation{Department of Physics and Astronomy, The University of Tennessee, Knoxville, Tennessee 37996, USA}
\author{Hao Zhang}
\affiliation{Department of Physics and Astronomy, The University of Tennessee, Knoxville, Tennessee 37996, USA}
\affiliation{Theoretical Division and CNLS, Los Alamos National Laboratory, Los
Alamos, New Mexico 87545, USA}
\author{Zoha Laraib}
\affiliation{Department of Physics and Astronomy, The University of Tennessee, Knoxville, Tennessee 37996, USA}
\author{Daniel M. Pajerowski}
\affiliation{Neutron Scattering Division, Oak Ridge National Laboratory, Oak Ridge, Tennessee 37831, USA}
\author{Kipton Barros}
\affiliation{Theoretical Division and CNLS, Los Alamos National Laboratory, Los
Alamos, New Mexico 87545, USA}
\author{Cristian~D.~Batista}
\affiliation{Department of Physics and Astronomy, The University of Tennessee, Knoxville, Tennessee 37996, USA}
\affiliation{Neutron Scattering Division, Oak Ridge National Laboratory, Oak Ridge, Tennessee 37831, USA}
\date{\today}
\begin{abstract}
We derive a renormalized classical spin (RCS) theory  for $S>1/2$ quantum magnets by constraining
a generalized classical theory that includes all multipolar fluctuations to a reduced  CP$^{1}$ phase space of dipolar SU(2) coherent states. 
When the spin Hamiltonian $\hat{\cal{H}}(S)$ is linear in the spin operators $\hat{\bm{S}}_j$ for each lattice site $j$, the RCS Hamiltonian $\tilde{\cal{H}}_{\rm cl}$ coincides with the usual classical model ${\cal H}_{\rm cl}= \lim_{S \to \infty} \hat{\cal H}(S)$. In the presence of non-linear terms,
however, the RCS theory is more  accurate than ${\cal H}_{\rm cl}$.
For the many  materials
 modeled by spin Hamiltonians with (non-linear) single-ion anisotropy terms,
the  use of the RCS theory is essential to accurately model  phase diagrams and to extract the correct Hamiltonian parameters from neutron scattering data.
\end{abstract}
\pacs{~}
\maketitle

The development of classical and semiclassical approximations to treat  quantum systems has played an important role in physics since the early days of quantum mechanics~\cite{schrodinger1926}. While semiclassical theories become exact when a ``control parameter'' $\alpha$ of the quantum mechanical Hamiltonian $\hat{\cal{H}}(\alpha)$ is sent to infinity,
${\cal H}_{\rm cl}= \lim_{\alpha \to \infty} \hat{\cal H}(\alpha)$, a vast amount of experimental evidence accumulated over many decades indicates that semiclassical approximations often remain  accurate for small values of $\alpha$. This 
is particularly evident in quantum magnetism, where $\hat{\cal H}$ is a
spin Hamiltonian and $\alpha$ is typically associated with the spin $S$ of the magnetic ions: ${\cal H}_{\rm cl}= \lim_{S \to \infty} \hat{\cal H}(S)$.
Although $S$ is of order one for most quantum magnets, semiclassical approaches 
are the standard tool to describe these materials because of their extraordinary success in reproducing experimental observations, such as the collective modes 
of magnetic materials.
Indeed, a very active area of quantum magnetism is the search for materials that exhibit strong deviations from semiclassical theories. While deviation is the rule for most quasi-1D magnets, finding examples of  quasi-2D and 3D magnets that fall into this category is challenging~\footnote{Large deviations from semiclassical behavior in 2D or 3D materials generally arise for low-$S$ materials, and near points where the classical spin Hamiltonian exhibits extensive ground state degeneracy}.

Given the importance of classical methods, we must observe
that the classical limit (CL) of a spin Hamiltonian $\hat{\cal H}$ is not unique, and the correct
choice of CL for a given system is crucial. In particular,
when $\hat{\cal H}$ contains nonlinear terms in the spin operators of a given site,
the traditional large-$S$ CL introduces systematic errors that can be avoided
by the use of an alternative limit. This has wide-ranging consequences. For example, 
Landau-Lifshitz (LL) dynamics and spin wave theory (SWT), both based on the large-$S$ limit,
are used extensively in the inverse modeling problem of extracting 
$\hat{\cal H}$ from scattering data.
These approaches underestimate the magnitudes of nonlinear contributions, such as
single-ion (SI) anisotropies or biquadratic interactions, by a substantial amount when $S$ is small.
Additionally,
many magnetic phenomena of current interest, such as skyrmions, are essentially
classical in nature, and classical methods are widely used to produce phase diagrams \cite{Leonov2015}.
We will show that these phase diagrams exhibit an $S$-dependence
that is lost in the large-$S$ CL. Finally, the choice of adequate
CL is essential as the starting point for developing semiclassical approaches, such 
as SWT and its generalizations.

Here we present an alternative CL appropriate for a wide range
of Hamiltonians, namely those for which bilinear (e.g., exchange) interactions are dominant,
but which also possess comparatively weak SI anisotropies or other terms that are nonlinear in
the spin operators. This alternative limit does not take $S$ as the control parameter but instead
uses $\lambda_1$, which labels the degenerate irreps of SU$(N)$. Correspondingly, quantum corrections are organized in powers of $1/\lambda_1$ instead of $1/S$, and the classical limit is obtained by sending $\lambda_1$ to infinity instead of $S$. The resulting Renormalized Classical Spin (RCS) Hamiltonian is identical to the traditional classical Hamiltonion
except that nonlinear terms have been renormalized by coefficients expressed in powers of $1/S$. These $1/S$ factors do not emerge from the introduction of higher-order quantum corrections; instead, they are the consequence of group theoretical considerations when comparing two different classical limits. Application of these renormalizations factors is straightforward and yields
a classical theory that precisely recovers the quantum expectation value of nonlinear terms with respect to any dipolar spin state.

{\it Renormalized Classical Theory.} 
Classical and semiclassical approximations are based on coherent states, which link the quantum and classical worlds.
The coherent states of a Lie group are obtained by applying 
the group elements to a reference state known as the highest weight state, and the resulting manifold of coherent states constitutes the phase space of the resulting classical theory~\cite{Perelomov72, yaffe1982large}. Quantum spin systems admit more than one CL since there 
is freedom to choose different Lie groups, each of which generates a different set of coherent states. A natural choice is the group of spin rotations, SU$(2)$, which leads to the 
traditional dipole-only CL of quantum spin systems. The resulting phase space of coherent states $|\Omega_j \rangle$ ($j$ is a lattice site index) is the 2D sphere, $S^2 \cong$ CP$^1$, 
which represents different possible orientations of dipole moments $\vec{\Omega}_j =\langle \Omega_j| \hat{\bm{S}}_j  |\Omega_j \rangle$ ($|\vec{\Omega}_j|=S$). 
The CL of the 
Hamiltonian $\hat{\cal{H}}$ is obtained by evaluating the expectation value of $\hat{\mathcal{H}}$ with respect to an arbitrary product of SU($2$) coherent states in the large-$S$ limit, i.e., by sending the spin irreps of SU$(2)$ to infinity,
\begin{equation}
{\cal H}_{\rm cl}[\vec{\Omega}]= \lim_{S \to \infty} \langle \vec{\Omega} | \hat{\cal H}(S) | \vec{\Omega} \rangle 
\label{eq:classlim}
\end{equation}
with $| \vec{\Omega} \rangle \equiv \bigotimes_j | \vec{\Omega}_j \rangle$.
The dynamics associated with the classical Hamiltonian are obtained by considering the Heisenberg equations of motion in the same limit, yielding 
\begin{equation}
\dot{\vec{\Omega}}_j =   \frac{\partial \cal{H}_{\rm cl}}{\partial \vec{\Omega}_j} \times {\vec{\Omega}}_j.
\label{eq:LL}
\end{equation}
This is the well-known LL equation~\cite{Landau35,Lakshmanan11}. Gilbert extended this dynamics with the introduction of damping in 1954~\cite{Gilbert54}, and the resulting LL-Gilbert (LLG) equation is now a fundamental tool in applied magnetism~\cite{Hillebrands02,Mattis98,Stiles06}. 

An alternative CL~\cite{Hao21, Dahlbom22, Dahlbom22b, RemundPohleAkagiShannon2022, PohleShannonMotome2023}, relevant when $S>1/2$, is obtained by considering coherent states of SU($N$) with $N=2S+1$: $\vert\Psi_j\rangle$ ~\footnote{
Note the use of SU$\left(N\right)$ coherent states does
not require an SU$(N)$-invariant Hamiltonian.  Instead, SU$(N)$ appears as the group
of possible time-evolution operators of an approximate local Hamiltonian. See SM.}. Just as an SU(2) coherent state may be uniquely associated with
a 3-vector of dipole moments, an SU($N$) coherent state may be uniquely associated
with a $(N^2-1)$-vector of dipole and multipole moments, $\Psi_j^\alpha = \langle \Psi_j \vert \hat{T}^\alpha \vert \Psi_j\rangle$, where $\hat{T}^\alpha$ are generators of SU($N$).  These generators may be selected such that $\hat{T}^\alpha = \hat{S}^\alpha$ for $\alpha=1\ldots3$.  To take the classical limit, the expectation value of $\hat{\mathcal{H}}$ is evaluated with respect to an arbitrary SU($N$) coherent state 
in the $\lambda_1\to\infty$ limit, where $\lambda_1$ labels irreps of SU($N$)~\footnote{$\lambda_1$ is the only finite eigenvalue when the maximal weight state is applied to the Cartan subalgebra. As with $S$ in the case of SU$(2)$, $\lambda_1$ is uniquely associated with the dimension of an irrep of SU$(N)$. See \cite{Gnutzmann98,Hao21} for details.},

\begin{equation}
{\cal H}_{\rm cl}^{{\rm SU}(N)}[\vec{\Psi}] = \lim_{\lambda_1 \to \infty}  
\langle \vec{\Psi} | \hat{\cal H}(\lambda_1) | \vec{\Psi} \rangle =
\langle \vec{\Psi} | \hat{\cal H} | \vec{\Psi} \rangle 
\label{eq:sunlim}
\end{equation}
with $| \vec{\Psi} \rangle \equiv \bigotimes_j | \vec{\Psi}_j \rangle$ . The final equality holds because the expectation value in the $\lambda_1\to\infty$ limit is the same as the expectation value in the fundamental represention \cite{SM}. In other words, the classical Hamiltonian is the \emph{exact} expectation value with respect to an SU($N$) coherent state. Note that $\hat{\mathcal{H}}$ retains its dependence on $S$, but $S$ is not used as the control parameter in the limiting procedure. The local phase space of this classical theory is CP$^{2S}$ rather than CP$^1$.

The set of SU$(2)$ coherent states, $|\vec{\Omega_j}\rangle$, is a submanifold of the set of SU$(N)$ coherent states, $|\vec{\Psi_j}\rangle$. This submanifold is simply the set of states obtained by applying group actions generated by the spin operators to the highest weight state while disregarding the additional generators in the algebra of SU$(N)$. It is therefore natural to define a constrained version of $\mathcal{H}^{{\rm SU}(N)}_{\rm cl}$ by limiting
the classical Hamiltonian obtained in the large-$\lambda_1$ limit to the coherent states of SU$(2)$:
\begin{equation}
{\cal \tilde{H}}_{\rm cl}[\vec{\Omega}] = \lim_{\lambda_1 \to \infty} \langle \vec{\Omega} | \hat{\cal H}(\lambda_1) | \vec{\Omega} \rangle = \langle \vec{\Omega} | \hat{\cal H}| \vec{\Omega} \rangle.
\label{eq:rcslim}
\end{equation}
This is will be referred to as the renormalized classical spin (RCS) Hamiltonian. 
 By construction, the phase space of ${\cal \tilde{H}}_{\rm cl}$ is the same as that of $\mathcal{H}_{\rm{cl}}$, namely the space of SU($2$) coherent states. Moreover, the associated dynamics will be that given in Eq.~\eqref{eq:LL}~\cite{SM}, with the substitution $\mathcal{H}_{\rm cl} \to \tilde{\mathcal{H}}_{\rm cl}$.~\footnote{The full, unconstrained SU$(N)$ theory \cite{Hao21} becomes relevant when anisotropies or other nonlinear terms are strong relative to the linear ones. In this case, the classical spin in the large-$\lambda_1$ limit (an SU(N) coherent state) no longer exhibits a rigid dipole and the restricted classical phase space of SU$(2)$ coherent states, the 2-sphere, no longer captures all the relevant degrees of freedom. In other words, both the traditional large-$S$ approach and the RCS theory become bad approximations.}

The significance of defining this Hamiltonian becomes clear when one considers the different behavior of the two limits on nonlinear terms.
In the $S \to \infty$ limit, an important simplification arises in the factorization of the expectation value of a product of on-site operators into the product of the expectation values of each individual operator:
\begin{eqnarray}
\lim_{S \to \infty} \langle \vec{\Omega} | \hat{S}^{\mu}_j  \hat{S}^{\nu}_j \dots  \hat{S}^{\eta}_j | \vec{\Omega} \rangle  &=& 
 \langle \vec{\Omega} |  \hat{S}^{\mu}_j | \vec{\Omega} \rangle   \langle \vec{\Omega} |   \hat{S}^{\nu}_j  | \vec{\Omega} \rangle \cdots 
\langle \vec{\Omega} |  \hat{S}^{\eta}_j | \vec{\Omega} \rangle
\nonumber \\
&=& \Omega^{\mu}_j \Omega^{\nu}_j \dots \Omega^{\eta}_j.
\label{eq:fac}
\end{eqnarray}

However, to describe a quantum mechanical system with a \emph{finite} value of $S$ that includes nonlinear terms in the components of an on-site spin operator $\hat{\bm{S}}_j$, Eq.~\eqref{eq:fac} becomes an extra approximation on top of the one made in Eq.~\eqref{eq:classlim}. We can obtain a more accurate classical approximation of $\hat{\cal H}(S)$ if we
we can avoid this step,
which is exact only in the $S \to \infty$ limit.

In contrast, Eq.~\eqref{eq:rcslim} shows that in the RCS theory the resulting Hamiltonian is exactly the quantum expectation value with respect to a finite-$S$ SU(2) coherent state. The two limits will differ precisely when the $S\to\infty$ limit demands application of the factorization rule

The RCS theory is simply the full SU($N$) theory constrained to the dipole sector.
It is applicable when the only collective modes of the Hamiltonian are dipole fluctuations,
which is typical when the exchange and Zeeman terms are strong relative to the nonlinear terms.
When this condition is satisfied, the full SU($N$) theory will predict
extra, non-dipolar modes that contribute little or no intensity to the dynamical spin structure factor. Moreover, these modes are overdamped (via loop corrections to the SU($N$) linear SWT) by the two-magnon continuum.
The RCS theory will not produce these spurious modes.
Further, the RCS Hamiltonian may be derived by applying a simple renormalization to ${\cal H}_{\rm cl}$.

\emph{Applying the RCS Theory} Consider
a broad class of spin Hamiltonians containing both interaction and single ion terms,
\begin{equation}
\hat{\mathcal{H}}=\underset{\textrm{Interaction}}{\underbrace{\hat{\mathcal{H}}^{\mathrm{Bil}}+\hat{\mathcal{H}}^{\mathrm{Biq}}}} + \underset{\textrm{SI}}{\underbrace{\hat{\mathcal{H}}^{\mathrm{Z}}+\hat{\mathcal{H}}^{\mathrm{A}}}}.
\label{eq:generic_hamiltonian}
\end{equation}
Each  term will be expressed
as polynomials of operators $\hat{\bm S}_j$ that may
represent  spin degrees of freedom for 
magnetic ions with weak spin-orbit interaction, or the total angular momentum $\hat{\bm J}_j$ for
magnetic ions with strong spin-orbit coupling.

The dominant interactions are typically bilinear,
\begin{equation}
    \hat{\cal{H}}^{\rm Bil} = \frac{1}{2} \sum_{i \neq j} \hat{S}^{\mu}_i {\cal J}^{\mu \nu}  \hat{S}^{\nu}_j.
\end{equation}

implying that they are linear in the spin operators of each site. Thus, taking the classical
limit -- both large-$S$ and RCS -- amounts to the substitution of each spin operator by a spin component: $\hat{S}^\mu_j \rightarrow \langle\Omega_j|S^\mu_j|\Omega_j\rangle = \Omega^\mu_j$. 
The same applies to the Zeeman term,
$
\hat{\cal{H}}^{\rm Z},
$
which is linear in $\hat{S}^\mu_j$.

Biquadratic interactions, $\hat{\mathcal{H}}^{\rm Biq}$, and SI anisotropies,
$\hat{\mathcal{H}}^{\rm A}$, require special treatment. First consider 
\begin{equation}
\hat{\cal{H}}^{\rm A}= \sum_{j, q, k} A_q^{(k)}  \hat{O}_{q}^{(k)}({\hat{\bm{S}}}_j),
\end{equation}
which includes only even powers of the spin operators due to time-reversal invariance. It is
expressed as a linear combination of
Stevens operators, $\hat{O}_{q}^{(k)}$, which span the $k$-irrep of SO(3).
The coefficients, $A_q^{(k)}$, are the crystal field parameters. 
The traditional CL of a Stevens operator is $\lim_{S\rightarrow\infty}\langle\Omega|\hat{O}_{q}^{(k)}|\Omega\rangle$, and the RCS limit is $\langle\Omega|\hat{O}_{q}^{(k)}|\Omega\rangle$, where the coherent state is given in the spin-$S$ representation with $S$ finite. Both
may be expressed as functions of the two angles that parameterize the SU$(2)$ coherent state $|\Omega\rangle$, and both will transform according to the same irrep. It follows that they are proportional.
In general we have
\begin{equation}
\tilde{\cal{H}}^{\rm A}_{\rm cl}= \sum_{j, q, k}  \tilde{A}_q^{(k)}  {\cal O}_{q}^{(k)}(\vec{\Omega}_j)
\end{equation}
with $\tilde{A}_q^{(k)} = c^{(k)} A_q^{(k)}$. The proportionality constants are~\cite{SM}, 
\begin{eqnarray}
c^{\left(2\right)} &=& 1-\frac{1}{2}S^{-1}, \quad
c^{\left(4\right)} = 1-3S^{-1}+\frac{11}{4}S^{-2}-\frac{3}{4}S^{-3}
\nonumber \\
c^{\left(6\right)} &=& 1-\frac{15}{2}S^{-1}+\frac{85}{4}S^{-2}-\frac{225}{8}S^{-3}+\frac{137}{8}S^{-4}-\frac{15}{4}S^{-5}
\nonumber
\end{eqnarray}
As expected, $\lim_{S \to \infty} \tilde{\cal{H}}^{\rm A}_{\rm cl} = {\cal{H}}^{\rm A}_{\rm cl}$ because $\lim_{S \to \infty} c^{(k)}=1$, but these renormalization factors become significant for low spin values. For instance, 
\begin{equation}
c^{\left(2\right)}(S=1)=1/2, \; c^{\left(4\right)}(S=2)=3/32, \;  c^{\left(6\right)}(S=3)=5/324. 
\label{eq:reno}
\end{equation}
Thus, if a spin-1 quadratic SI anisotropy is approximated classically taking the $S \to \infty$ limit, the anisotropy strength will be 
underestimated by a factor $1/c^{(2)} = 2$.  Errors become even more severe for higher-order SI anisotropies.

The remaining term in Eq.~\eqref{eq:generic_hamiltonian} corresponds to biquadratic or higher-order 
interactions, which may be significant in Mott insulators that are not deep inside the Mott regime or in $f$-electron magnets where the spin-orbit coupling is comparable to the intra-atomic Coulomb interaction.
As a simple illustration, consider the isotropic biquadratic interaction,

\begin{equation}
\hat{\cal{H}}^{\rm Biq} = \frac{1}{2} \sum_{i \neq j} {\cal K}_{ij} (\hat{\bm{S}}_i \cdot  \hat{\bm{S}}_j)^2.
\end{equation}
After evaluating $\hat{\cal{H}}^{\rm Biq}$ in  the $S\rightarrow\infty$ limit and comparing the result
to its expectation value with respect to a finite-$S$ coherent state~\cite{SM}, the RCS Hamiltonian
is found to be
\begin{equation}
\tilde{\cal{H}}^{\rm Biq}_{\rm cl}= \frac{1}{2} \sum_{ i \neq j} {\cal K}_{ij} \left[ r (\vec{\Omega}_1 \cdot \vec{\Omega}_2)^2 - \frac{1}{2} \vec{\Omega}_1 \cdot \vec{\Omega}_2   + S^3+ \frac{S^2}{4} \right]
\end{equation}
with
$
r = \left( 1 - \frac{1}{S} + \frac{1}{4S^2} \right). 
$
Besides being renormalized, the biquadratic interaction generates a bilinear term that is absent in the large-$S$ limit and
is comparable in amplitude to the biquadratic interaction for $S=1$. Moreover, the renormalization factor becomes $r(S=1)=1/4$, implying that the amplitude of the biquadratic term is 4 times smaller than $r(S\to \infty)=1$.

\begin{figure}
\centering
\includegraphics[width=1.0\columnwidth]{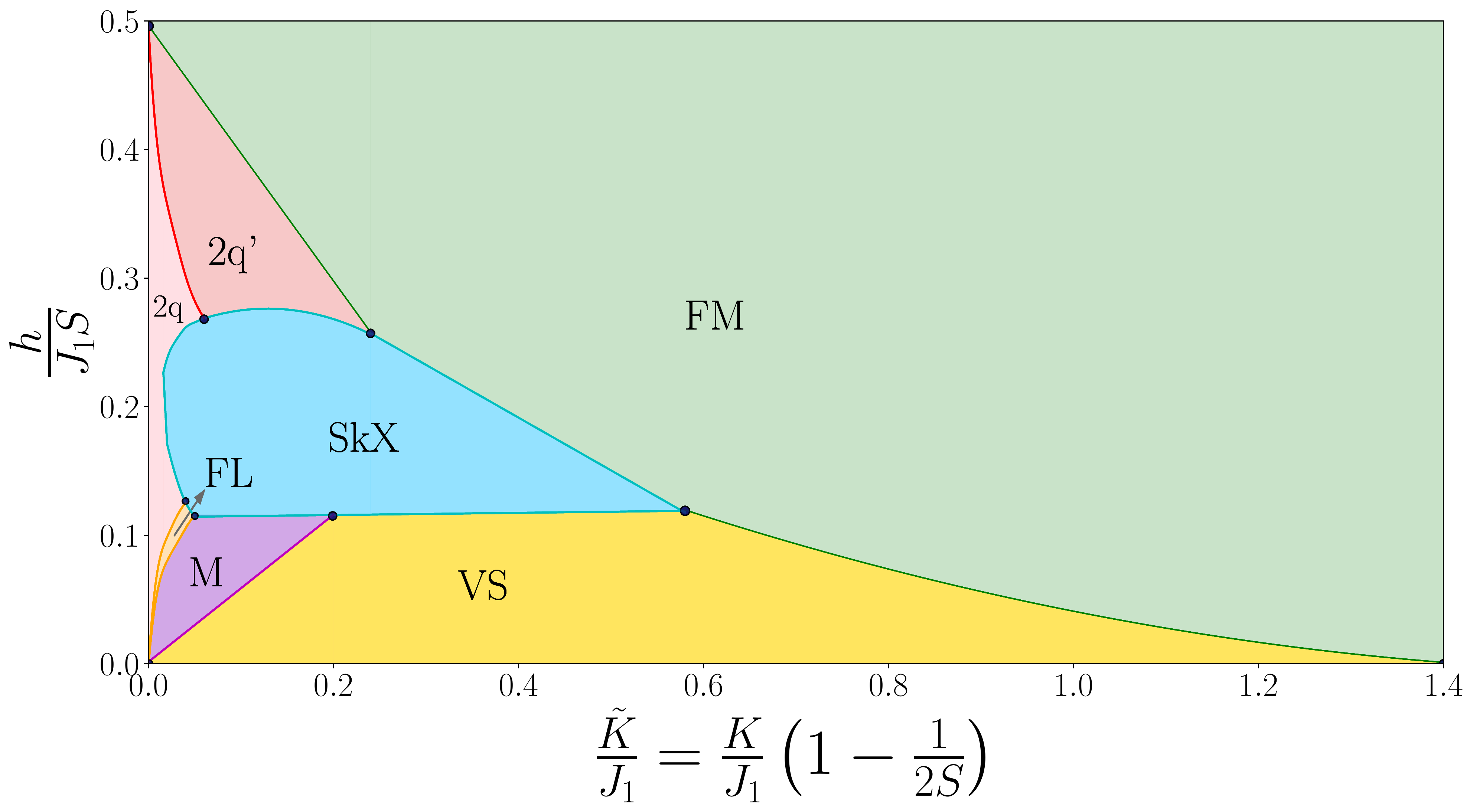}
\vspace{-0.7cm}
\caption{Variational phase diagram of the spin Hamiltonian Eq.~\eqref{eq:spin-model}.  Phase notation and additional details may be found in reference~\citep{Leonov2015}.}
\label{Fig:fig1}
\end{figure}

{\it $S$-dependent Phase Diagrams.} If ${\cal H}_{\rm cl} \neq \tilde{\cal H}_{\rm cl}$, the RCS Hamiltonian will produce an $S$-dependent thermodynamic phase diagram that coincides with the usual classical phase diagram only in the $S\to \infty$ limit. This observation has many implications for real magnets, which we illustrate with a model that has been extensively studied in the context of magnetic skyrmion crystals~\citep{Leonov2015}. 
The spin-$S$ Hamiltonian, 
\begin{equation}
    \hat{\mathcal{H}} = \frac{1}{2} \sum_{i \neq j} J_{ij} \hat{\bm{S}}_i \cdot \hat{\bm{S}}_j
    - h \sum_i \hat{S}_i^z - \frac{K}{2} \sum_i (\hat{S}_i^z)^2,
    \label{eq:spin-model}
\end{equation}
includes ferromagnetic nearest-neighbor exchange interactions and antiferromagnetic next-nearest-neighbor interactions on a triangular lattice: $J_{ij} =J_1 < 0$ ($J_{ij} =J_2 > 0$) for nearest (next-nearest) neighbors spins $i$ and $j$ and $J_{ij}=0$ otherwise. The second and third terms are, respectively, Zeeman coupling to a magnetic field along the $z$-axis and easy-axis ($K>0$) SI anisotropy. Leonov and Mostovoy~\citep{Leonov2015} used a variational scheme to compute  the classical phase diagram of this model, shown in Fig.~\ref{Fig:fig1}, by taking the large-$S$ limit, i.e., by using the classical Hamiltonian ${\cal H}_{\rm cl}$.  For sufficiently small magnetic fields, the lowest energy state of $\mathcal{H}_{\rm cl}$ is the ``vertical spiral'' (VS) phase with  a polarization plane parallel to the $z$-axis and a propagation wave vector  $\pm {\bm Q_{\nu}}$ ($\nu =1,2,3$) parallel to the three possible directions  related by 120$^{\circ}$ rotations about the $z$-axis. 
Several multi-${\bm Q}$ orderings corresponding to a superposition of more than one spiral are induced upon increasing $h$, as described in detail in Ref.~\cite{Leonov2015}.
The most interesting phase is the triple-${\bm Q}$ skyrmion crystal (SkX) that extends over the interval $ 0.05 \lesssim K/|J_1| \lesssim 0.6 $ 
However, if we use the RCS theory,
\begin{equation}
    \tilde{\mathcal{H}}_{\rm cl} = \frac{1}{2} \sum_{i \neq j} J_{ij} {\vec{\Omega}}_i \cdot {\vec{\Omega}}_j
    - h \sum_i {\Omega}_i^z - \frac{\tilde{K}(S)}{2} \sum_i (\hat{\Omega}_i^z)^2,
    \label{eq:cl-spin-model}
\end{equation}
with $\tilde{K} = K (1 - \frac{1}{2S})$, the range of stability of the SkX phase is $S$-dependent. In particular, for $S=1$ the SkX phase is stable over a range of $K$ values that is twice as large as the range obtained for $S \to \infty$, i.e., lowering the spin of the magnetic ions becomes a guiding principle 
to find more robust SkXs in centrosymmetric materials. It is also noteworthy that exactly the same 
phase diagram is obtained for $S=1$ 
if we use direct products of coherent states of SU(3) as a variational space~\cite{Hao21}. Finally, note that the only difference between $\tilde{\mathcal{H}}_{\rm cl}$ and $\mathcal{H}_{\rm cl}$ is a renormalization of the SI anisotropy. More drastic differences between the  phase diagrams of the two classical models, such as the presence of different phases, can be expected if more than one Hamiltonian parameter is renormalized.

{\it More Accurate Dynamics.} Besides producing different phase diagrams, $\tilde{\mathcal{H}}_{\rm cl}$ and $\mathcal{H}_{\rm cl}$ will in general lead to different dispersion relations of the normal modes (spin-waves) of a given phase. The linear SWT (LSWT) is obtained by quantizing the harmonic oscillators of each normal mode: the spin-waves of the CL become magnons of the quantum mechanical theory. Since Hamiltonian parameters are typically extracted by fitting the magnons measured with INS, using the renormalized classical Hamiltonian is critical to extract correct values. 
Consider the fully polarized phase of the Hamiltonian Eq.~\eqref{eq:cl-spin-model}. This has the \emph{exact} single-magnon dispersion
\begin{equation}
\omega_{\bm k}= S [{\cal J}({\bm k})- {\cal J}({\bm 0})] +h + \tilde{K}S,
\label{eq:polarized}
\end{equation}
 where
${\cal J}({\bm k})= \sum_j e^{i {\bm k} \cdot {\bm r}_{ij}} J_{ij}$. This coincides with the dispersion relation that is obtained from  $\tilde{\mathcal{H}}_{\rm cl}$, but differs from the result $\omega_{\bm k}=S [{\cal J}({\bm k})-{\cal J}({\bm 0})]+h+ KS$ that is obtained in the large-$S$ limit. 
In other words, the value of the actual SI anisotropy is $(1-1/2S)^{-1}$  times bigger than the value obtained by fitting the magnon dispersion with the unrenormalized LSWT, giving a relative correction of 100$\%$ for $S=1$ and still $17\%$ for $S=7/2$. Moreover, from
Eq.~\eqref{eq:reno}, in the presence of quartic ($q=4$) and sixth ($q=6$) order SI anisotropy terms, the unrenormalized LSWT predicts amplitudes that are of order $10$ and $100$ times bigger than the actual values for the lowest values of $S$ compatible with these anisotropies. 

Progress in quantum magnetism relies heavily on inelastic neutron scattering (INS) data. The recent development of performant 
LSWT codes (e.g., SpinW) has been instrumental in opening the bottleneck between observables and model Hamiltonians,  providing crucial information closer to the beginning of the materials life cycle
\cite{SpinW_Toth_2015}. Nearly~20\% of reports from the last year that make reference to SpinW include Hamiltonians with nonlinear terms in site spin operators.  
More accurate values may be obtained using either the renormalization factors  $c^{(k)}$ or the recipe for generation of non-tabulated terms. Looking forward, estimates should be made either using the appropriate SU($N$) LSWT or 
the renormalized SU(2) LSWT to exctact more accurate models.
The open-source code Sunny provides an implementation of the RCS theory \footnote{{h}ttps://github.com/SunnySuite/Sunny.jl}.


Previous works have used Bose operator expansions to  renormalize SU($2$) LSWT for spin Hamiltonians with SI anisotropy terms. In 1961, Oguchi and Honma performed a $1/S$ expansion of the square root that appears in the Holstein-Primakoff transformation to order $1/S^0$ and in this way derived the $(1-1/2S)$ renormalizalization factor of the quadratic SI anisotropy~\cite{Oguchi60}. 
In 1976, Kowalska and Lindg\r{a}rd found the same renormalization using a generalized crystal-field Hamiltonian with an expansion parameter of the crystal field strength divided by the exchange field strength~\cite{Kowalska76}.  Later efforts have echoed these results, whether using Holstein-Primakoff~\cite{Rastelli79} or Dyson-Maleev~\cite{Rastelli84} transformations.  However, since these works focused on the linearized dynamics (LSWT), the renormalization of the underlying classical theory and the corresponding non-linear LL equations were not discussed. 
What remained hidden is that the renormalization of the linearized LSWT Hamiltonian arises from a more fundamental renormalization of the classical Hamiltonian. The failure to recognize this fact
presents a danger when adding only some $1/S$ corrections to LSWT calculations.
Specifically, the renormalization presented here can change the \emph{classical} ground state
that is used as the starting point for a spin-wave calculation. If $1/S$ corrections
are added to the LSWT calculation only, and not to the calculation of the classical ground
state, LSWT dispersion calculations may fail, either by gapping out Goldstone modes or by predicting unphysical, imaginary frequencies.

Recall that $\tilde{\cal{H}}_{\rm{cl}}$ is obtained by restricting the generalized dynamics of SU$(2S+1)$ coherent states (phase space CP$^{2S}$) to the submanifold of dipoles (SU(2) coherent states in CP$^1$) Since  CP$^{2S}$ contains $2S$ pairs of conjugate coordinates and momenta, the number of normal modes per spin is $2S$. The quantization of these normal modes is implemented by introducing Holstein-Primakoff bosons $b_{j \nu}$ with  $2S$ flavors ($1 \leq \nu \leq 2S$). The restriction of the linearized dynamics to the plane tangent to the the CP$^{1}$ phase space of SU(2) coherent states is equivalent to projecting out 
the $2S-1$ bosons representing fluctuations that are orthogonal to this plane.

Quantum corrections to the generalized spin-wave Hamiltonian are implemented via a loop expansion for the propagators of the $2S$ bosons (one for each flavor). This loop expansion is actually an expansion in  $1/\lambda_1$, where $\lambda_1$ labels degenerate irreps of SU($2S+1$) (generalization of the $1/S$ expansion)~\cite{Do21, Hao21}. The restricted dynamics is obtained by keeping only the bosonic propagator for the  magnon modes and integrating out the remaining modes.~\footnote{The internal lines of the diagrammatic expansion of this propagator can still have any flavor} 

These observations reveal that the RCS Hamiltonian and the corresponding renormalized SWT (including quantum corrections) is a projection of a generalized SWT, the ``control parameter'' of which is $\lambda_1$ instead of $S$. Thus, it is not surprising that the renormalization factors contain higher order corrections in $1/S$. The most important corollary is that the renormalized SWT with the above-mentioned quantum corrections still preserves the Goldstone modes of theories that spontaneously break continuous symmetries because these symmetries are preserved to each order of the $1/\lambda_1$ expansion. 

In summary, we have provided a general renormalization procedure which is essential for deriving accurate classical and  semiclassical approximations of spin Hamiltonians with nonlinear terms. While the procedure was illustrated with the most common SI anisotropy terms and isotropic biquadratic interactions, more general cases, such as anisotropic four-spin interactions, can naturally appear in materials with strong spin-orbit coupling, such as $4f$-electron systems. Since semi-classical methods are the most common approximation used for solving the inverse scattering problem, the renormalization procedure presented here is necessary for extracting correct Hamiltonian parameters from INS data. In particular, the traditional method of fitting the measured INS data with 
LSWT (large-$S$ limit) can lead to serious inconsistencies between the neutron-derived parameters and those derived from other measurements.

\emph{Acknowledgements} We thank Martin Mourigal, Alan Tennant and Xiaojian Bai for helpful discussions.
This work was funded  by the U.S. Department of Energy, Office of Science, 631
Office of Basic Energy Sciences, under Award No. DE-SC-DE-SC-0018660.  K.~B. and H.~Z. acknowledge support from the LANL LDRD program.  Z.~L. (phase diagram of Fig.~\ref{Fig:fig1}) acknowledges support from U.S. Department of Energy, Office of Science, Office of Basic Energy Sciences, Materials Science and Engineering Division.  D.~P. acknowledges support by the DOE Office of Science (Office of Basic Energy Sciences).  
This research used resources of the Oak Ridge Leadership Computing Facility at the Oak Ridge National Laboratory, which is supported by the Office of Science of the U.S. Department of Energy under Contract No.~DE-AC05-00OR22725. 
This manuscript has been authored by UT-Batelle, LLC, under contract DEAC05-00OR22725 with the US Department of Energy (DOE). The US government retains and the publisher, by accepting the article for publication, acknowledges that the US government retains a nonexclusive, paid-up, irrevocable, worldwide license to publish or reproduce the published form of this manuscript, or allow others to do so, for US government purposes. DOE will provide public access to these results of federally sponsored research in accordance with the DOE Public Access Plan (http://energy.gov/downloads/doepublic-access-plan).

\bibliographystyle{apsrev4-2}
\bibliography{ref}

\end{document}


\title{Supplemental Material for Renormalized Classical Theory of Quantum Magnets}
\author{David Dahlbom}
\affiliation{Department of Physics and Astronomy, The University of Tennessee, Knoxville, Tennessee 37996, USA}
\author{Hao Zhang}
\affiliation{Department of Physics and Astronomy, The University of Tennessee, Knoxville, Tennessee 37996, USA}
\author{Zoha Laraib}
\affiliation{Department of Physics and Astronomy, The University of Tennessee, Knoxville, Tennessee 37996, USA}
\author{Daniel M. Pajerowski}
\affiliation{Neutron Scattering Division, Oak Ridge National Laboratory, Oak Ridge, Tennessee 37831, USA}
\author{Kipton Barros}
\affiliation{Theoretical Division and CNLS, Los Alamos National Laboratory, Los
Alamos, New Mexico 87545, USA}
\author{Cristian~D.~Batista}
\affiliation{Department of Physics and Astronomy, The University of Tennessee, Knoxville, Tennessee 37996, USA}
\affiliation{Neutron Scattering Division, Oak Ridge National Laboratory, Oak Ridge, Tennessee 37831, USA}
\date{\today}
\maketitle

In this supplement, we review
some relevant facts about generalized coherent states; derive
the Landau-Lifshitz (LL) equations using coherent states of SU$(2)$; and generalize these 
equations using coherent states of SU$(N)$. We next
constrain the generalized dynamics to a dynamics of dipoles only. The resulting
equations of motion will have the same form as the LL equations
but nonlinear terms will be renormalized. Readers interested in the
renormalization factors alone may skip directly to Sections \ref{sec:renorm_on-site_anisotropies} and \ref{sec:renorm_biquadratic}.

As a precaution, 
we emphasize that 
the use of SU$(N)$ coherent states does
not require an SU$(N)$-invariant Hamiltonian. Indeed, it is very 
often the terms that break continuous symmetries,
such as on-site anisotropies, that motivate the use of the techniques
presented here. Instead, SU$(N)$ appears as the group
of possible time-evolution operators of an approximate local Hamiltonian.
SU$(N)$ in particular is the relevant group simply because it is the
set of all possible transformations that may be applied to an $N$-level
quantum system.

\subsection{Generalized coherent states \label{sec:generalized_cs}}

A generalized coherent state of a particular group may be realized
by applying any action of this group to a reference state \cite{Perelomov72, gilmore1972geometry}.
For SU$(2)$, the reference state is  
taken to be the $z$-polarized state $|S\rangle$:  ${\hat S}^z |S \rangle = S |S \rangle$.
Group actions may be generated by exponentiating linear combinations
of the spin operators, which form a basis for the Lie algebra $\mathfrak{su}(2)$.
Thus an arbitrary SU$(2)$ coherent state may be written,
\begin{equation}
\left|\Omega\right\rangle =e^{i\sum_{\alpha=1}^{3}c_{\alpha}\hat{S}^{\alpha}}\left|S\right\rangle, \label{eq:arbitrary_SU(2)_cs}
\end{equation}
for real coefficients $c_{\alpha}$. The set of all coherent states
of a group is generated by considering the application of all possible
group actions to the reference state. When considering the $S=\frac{1}{2}$
representation of SU$(2)$, this procedure is simply a
recipe for producing any pure state.

Different group actions may produce equivalent coherent states, i.e.,
states differing only by an overall phase factor~\cite{Perelomov72}. When this is true, the manifold
of coherent states will have lower dimensionality than the group
that was used to generate the states. For example, a general group
element of SU$(2)$ may be parameterized by three numbers,
such as the $c_{\alpha}$ above or the more familiar Euler angles,
but the manifold of coherent states is only 2-dimensional. This fact
is made more obvious after setting up a one-to-one correspondence
between coherent states and a real 3-vector through
a map that eliminates overall phase:
\begin{equation}
\left|\Omega\right\rangle \mapsto \left\langle \Omega\right|\hat{S}^{\alpha}\left|\Omega\right\rangle \equiv\Omega^{\alpha},
\label{eq:omega_vector}
\end{equation}
where $\alpha=x,y,z$. This map, which corresponds to the familiar
Bloch sphere construction, associates to each distinct coherent state a point
on the sphere. If phase-equivalence is treated appropriately, the map is invertible,
so the correspondence between points on a sphere and
coherent states of SU$(2)$ is one-to-one \footnote{
The inverse mapping may be constructed as follows. 
When constructing conherent states, take the highest-weight state as the reference state.
(For details on the highest-weight state in this context, see \cite{Hao21}.)
Construct the operator $\mathfrak{n} = \Psi^\alpha \hat{T}^\alpha$,
summing over $\alpha$. The highest-weight eigenvector of $\mathfrak{n}$ is then the complex-vector representation 
of the coherent state corresponding to $\Psi^\alpha$. For more details, see Part III of the supplementary material to \cite{zhang2022CP2Skyrmions}.}.
Consequently, we may parameterize the set of coherent states using only two angles.
In particular, if we choose to represent a coherent state as in Eq.~\eqref{eq:arbitrary_SU(2)_cs}, we may write
the three coefficients $c_{\alpha}$ as functions of these two angles. The choice
of a particular parameterization involves fixing a gauge.

We emphasize these points because the manifold of coherent states
will become the classical phase space in the appropriate limit \cite{yaffe1982large, Hao21}. Thus
the choice of group used to define coherent states is also a choice
about the structure of the resulting classical dynamics, which will have a local Hamiltonian structure with $N-1$ degrees of freedom.
For example, the choice $N=2$ yields the Landau-Lifshitz dynamics, which describes the evolution of a dipole on the sphere; this 2-dimensional manifold can be locally interpreted as a canonical phase space by the Darboux-Lie theorem \cite{hairer2013geometric}.

The construction of coherent states of SU$(N)$ proceeds
in the same manner. Group actions may now be generated by exponentiating
linear combinations of the elements of the Lie algebra $\mathfrak{su}(N)$.
This algebra is $(\text{\ensuremath{N^{2}}-1})$-dimensional,
and is thus spanned by $N^{2}-1$ generators, $\hat{T}^{\alpha}$. The generalization of Eq.~\eqref{eq:arbitrary_SU(2)_cs} to an arbitrary
SU$(N)$ coherent state, $\left|\Psi\right\rangle $, is
\begin{equation}
\left|\Psi\right\rangle =e^{i\sum_{\alpha=1}^{N^{2}-1}c_{\alpha} \hat{T}^{\alpha}}\left|S\right\rangle,\label{eq:arbitrary_su(N)_cs}
\end{equation}
for real coefficients $c_\alpha$. We continue to take $|S\rangle$ as the reference state. In the fundamental ($N$-dimensional)
representation,
this is simply a recipe for generating all pure states. As before,
it is possible to set up a one-to-one correspondence between 
coherent states and a vector of expectation values:
\begin{equation}
\left|\Psi\right\rangle \mapsto\left\langle \Psi\right|\hat{T}^{\alpha}\left|\Psi\right\rangle \equiv\Psi^{\alpha}.\label{eq:psi_vectors}
\end{equation}
The number of components in this vector is the dimension of
the group. For SU$(N)$, this is $N^{2}-1$. The resulting
manifold of  coherent states is again a lower-dimensional object than the 
group itself.
Specifically, it is the complex projective space CP$^{N-1}$ \cite{bengtsson2017geometry}.
(Note that CP$^{1}$ is topologically equivalent to the two-sphere.) This is a $2\left(N-1\right)$-dimensional
manifold, and we may in principle parameterize any SU$(N)$
coherent state with $2(N-1)$ angles. An important consequence 
is that the $N^2-1$ expectation values, $\Psi^{\alpha}$, are constrained,
just as the three expectation values in Eq.~\eqref{eq:omega_vector} are
restricted to the sphere.

Without loss of generality, it is always possible to choose the generators
$\hat{T}^{\alpha}$ such that the first three correspond to the spin operators:

\begin{equation}
\hat{T}^{\alpha}=\hat{S}^{\alpha},\quad\alpha=1,2,3.\label{eq:T1-3_as_S1-3}
\end{equation}
With such a choice, the first three generators form an $\mathfrak{su}(2)$
subalgebra of $\mathfrak{su}(N)$. The manifold of coherent
states generated by this subalgebra,
\begin{equation}
\left|\Psi\right\rangle =e^{i\sum_{\alpha=1}^{3}c_{\alpha}\hat{T}^{\alpha}}\left|S\right\rangle,
\label{eq:arbitrary_restricted_su2_cs}
\end{equation}
is identical the one defined in Eq. (\ref{eq:arbitrary_SU(2)_cs}).
The resulting coherent states may therefore be labeled $\left|\Omega\right\rangle $
as well. The distinction is that this manifold of $SU\text{\ensuremath{\left(2\right)}}$
coherent states is a submanifold of the larger manifold of SU$(N)$
coherent states. The latter may be obtained by considering group actions
generated by the entire set of generators $\hat{T}^{\alpha}$, not just the first three.

To make this discussion more concrete, consider a spin-$S$ Hamiltonian. For such a Hamiltonian,
it will be natural to consider the spin-$S$ representation of the spin operators, i.e., the conventional representation
that acts an $N=2S+1$ dimensional Hilbert space. In this case, the SU(2) coherent states
may be parameterized explicitly by using Eq.~\eqref{eq:arbitrary_SU(2)_cs} with appropriate choice of gauge.
The result of the calculation is,

\begin{equation}
    \left\vert \Omega \right\rangle = \sqrt{\left(2S\right)!} \sum_m \frac{u^{S+m}v^{S-m}}{\sqrt{ \left( S+m \right)!\left( S-m \right)! }}\left\vert S, m\right\rangle, \label{eq:cs_parameterization}
\end{equation}
where $u\left(\theta,\phi\right)=\cos\left(\theta/2\right)e^{i\phi/2}$ and $v\left(\theta,\phi\right)=\sin\left(\theta/2\right)e^{-i\phi/2}$ \cite{auerbach1994}. An important observation is that, when $S=1/2$, this formula gives a complete parameterization of all the pure states of a two-level system. Physically, these are all the points on a Bloch sphere or, equivalently, the set of all possible states of a dipole of fixed magnitude. When $S>1/2$, the SU(2) coherent states parameterized by this formula are only a submanifold of the set of all pure states of an $N=2S+1$ system. The set of all of the pure states of an $N$-level system do however correspond in a one-to-one fashion with the coherent states of SU($N$). These may be explicitly parameterized in a number of ways (see \cite{nemoto2000generalized} for one approach).  Physically, 
this larger set of states corresponds to all spin states of a spin-$S$, which includes states with reduced or zero dipole moment as well as multi-polar components. Note, however, that the SU(2) coherent states described above, which described a dipole of fixed-length, always exist as a submanifold within this larger set of coherent states. The importance of 
considering dipolar SU(2) coherent states as embedded in all of the coherent states of SU(N) will become important when we consider the classical limit.

We note that, in principle, there are many formal embeddings SU$(2)$ coherent states (not necessarily dipolar in character)
in the manifold of SU$(N)$ coherent states. In the following, references to $SU(2)$ coherent states
as part of the larger manifold of SU$(N)$ coherent
states will always be made specifically to those states generated by the 
spin operators, as in Eq.~\eqref{eq:arbitrary_restricted_su2_cs}. In other words, we will only be concerned with that set of SU(2) coherent states that correspond to dipolar states.

As an initial motivation for considering this larger set of coherent states, 
consider the single-site Hamiltonian containing only a quadrupolar term:
\begin{equation}
\hat{\cal{H}}=\left(\hat{S}^{z}\right)^{2}.
\label{eq:hamiltonian_quad_aniso}
\end{equation}
The time-evolution operator will simply be
\[
U\left(t\right)=e^{i\left(\hat{S}^{z}\right)^{2}t},
\]
where we have set $\hbar=1$. Note that this 
does not generate an SU$(2)$ group action, since $(\hat{S}^{z})^{2}$
does not belong to the Lie algebra $\mathfrak{su}(2)$. Put
more concretely, there is no way to write $(\hat{S}^{z})^{2}$
as a linear combination of the spin operators. When the time evolution
operator is applied to some arbitrary state, say the $z$-polarized
state,
\[
e^{i\left(\hat{S}^{z}\right)^{2}t}\left|S\right\rangle ,
\]
it will typically generate motion away from the two-sphere generated
by the spin operators. This presents
a problem for a classical dynamics defined on this sphere. In contrast, $(\hat{S}^{z})^{2}$
can always be written as a linear combination of generators of SU$(N)$ with $N>2$. 
 The operator therefore belongs to $\mathfrak{su}(N)$, and the
time evolution operator will generate a coherent state of
SU$(N)$. Such a state will always be representable in the classical
limit taken with respect to SU$(N)$ coherent states. 
This geometric observation has analytical consequences that will be
explored in the next section.

\subsection{The classical limit taken with respect to coherent states \label{sec:classical_limit}}

We briefly review the procedure for taking a classical limit with
respect to coherent states of SU$(2)$ and SU$(N)$.
Details are available in \cite{Hao21}. For simplicity, we will only consider
expressions that are polynomials in spin operators, which covers all
Hamiltonians of interest here.

A classical limit may be derived by replacing operators with expectation
values, where these expectation values are calculated with respect
to the coherent states of a group in the limit where the representation
of this group is sent to infinity. The label of the representation serves
as the control parameter in this limiting procedure. For example, the
limit taken with respect to coherent states of SU$(2)$ uses $S$
as the control parameter.

An expression such as
\[
\lim_{S\rightarrow\infty}\left\langle S\right|\hat{S}^{z}\left|S\right\rangle =\lim_{S\rightarrow\infty}S
\]
becomes infinite in the infinite-$S$ limit, so it is formally necessary to renormalize after
taking the limit. Typically the expression is simply renormalized
to $S$, where $S$ refers to the angular momentum of the original quantum Hamiltonian
\footnote{Alternatively, one may consider taking $\hbar\rightarrow0$
together with $S\rightarrow\infty$ while holding $\hbar S$ constant.}. 
In the remainder of this note, the action of taking the limit will
always be assumed to also include this renormalization step. As a consequence,
for simple terms that are linear in the spin operators, the following holds
\[
\lim_{S\rightarrow\infty}\left\langle \Omega\right|\hat{S}^{\alpha}\left|\Omega\right\rangle =\left\langle \Omega\right|\hat{S}^{\alpha}\left|\Omega\right\rangle.
\]
In other words, the expectation value taken in the fundamental representation
is the same (after renormalization) in infinite-$S$ representation.
This gives us the substitution rule,
\begin{equation}
\hat{S}^{\alpha}\rightarrow\left\langle \Omega\right|\hat{S}^{\alpha}\left|\Omega\right\rangle =\Omega^{\alpha}, \label{eq:su2_basic_sub_rule}
\end{equation}
for spin operators when they appear as isolated terms.

A subtlety arises when considering the expectation of a product of
operators, e.g., $\hat{S}^{\alpha}\hat{S}^{\beta}$. It can be shown,
\begin{equation}
\left\langle \Omega\right|\hat{S}^{\alpha}\hat{S}^{\beta}\left|\Omega\right\rangle =\left\langle \Omega\right|\hat{S}^{\alpha}\left|\Omega\right\rangle \left\langle \Omega\right|\hat{S}^{\beta}\left|\Omega\right\rangle +\mathcal{O}\left(S\right),\label{eq:expectation_of_prod_with_corrections}
\end{equation}
that is, the expectation of the product of operators is the product
of the expectations plus corrections terms of order $S$ or lower.
Since the first term is of order $S^2$, the remaining terms will disappear
in the large-$S$ limit:
\[
\lim_{S\rightarrow\infty} \left\langle \Omega\right|\hat{S}^{\alpha}\hat{S}^{\beta}\left|\Omega\right\rangle =\left\langle \Omega\right|\hat{S}^{\alpha}\left|\Omega\right\rangle \left\langle \Omega\right|\hat{S}^{\beta}\left|\Omega\right\rangle .
\]
This gives us the factorization rule,
\begin{equation}
\hat{S}^{\alpha}\hat{S}^{\beta}\rightarrow\left\langle \Omega\right|\hat{S}^{\alpha}\left|\Omega\right\rangle \left\langle \Omega\right|\hat{S}^{\beta}\left|\Omega\right\rangle =\Omega^{\alpha}\Omega^{\beta},\label{eq:factorization_rule}
\end{equation}
which may be generalized to an arbitrary product of operators as in Eq.~(8) of the main paper.
The factorization rule is an approximation and only becomes
exact in the infinite representation limit. An important motivation for
introducing SU$(N)$ coherent states is the elimination of this
approximation.

A final point should be made about taking the classical limit of more complicated
expressions containing groups of terms modified by common coefficients. If the
relative value of these coefficients is to be maintained after taking the limit,
only terms of highest-order in $S$ within each group will survive. For example,
consider the single-ion Hamiltonian,
\begin{equation}
    \hat{\mathcal{H}} = A\left(\hat{S}^z\right)^2  + D\left( \left(\hat{S}^x\right)^2 + \left(\hat{S}^y\right)^2 + \left(\hat{S}^z\right)^4 \right).
\end{equation}
We may replace $A$ with $A/S^2$ and $D/S^4$ to maintain
the relative importance of the two contributions in the large-$S$ limit.
This will have the effect of killing terms of order less than $S^2$ in the $A$
group and less than $S^4$ in the $D$ group. Thus,
\begin{equation}
    \lim_{S\rightarrow\infty}\hat{\mathcal{H}} = A\left(\Omega^z\right)^2 + D\left(\Omega^z\right)^4.
\end{equation}

The procedure for deriving the classical limit with SU$(N)$ coherent states
is essentially identical to the above. However, since SU$(2)$ is no 
longer the group being used to generate coherent states,
$S$, which labels irreps of this group, can no longer be used as the
control parameter. 
We instead consider degenerate irreps of SU$(N)$. These are labeled by a single
number, $\lambda^{(N)}_1$, the highest weight eigenvalue, and the classical limit will correspond to sending $\lambda^{(N)}_1\rightarrow\infty$. We note that the meaning of this parameter depends on $N$, hence the superscript, and typically we will consider $\lambda_1^{(2S+1)}$ when taking the limit of a spin-$S$ system.
The classical limit
is again determined by replacing operators with expectation
values, where these expectation values are now be calculated with
respect to SU$(N)$ coherent states in the $\lambda^{(N)}_{1}\rightarrow\infty$
limit. See \cite{Hao21} for details. 

Since we are working with coherent states of SU$(N)$,
we are free to express any operator in terms of the Lie algebra $\mathfrak{su}(N)$.
The latter may be thought of as the space of $N\times N$ Hermitian
matrices, with generators $\hat{T}^{\alpha}$ serving as a basis for this
space. For each of these generators we will have
\begin{equation}
\lim_{\lambda^N_1\rightarrow\infty} \left\langle \Psi\right|\hat{T}^{\alpha}\left|\Psi\right\rangle =\left\langle \Psi\right|\hat{T}^{\alpha}\left|\Psi\right\rangle =\Psi^{\alpha}\label{eq:sun_sub_rule}.
\end{equation}
That is, the expectation value in the infinite-$\lambda_{1}$ limit
(with appropriate renormalization) is the same as the expectation
value in the fundamental representation. Moreover, since the $\hat{T}^{\alpha}$
serve as a basis for all Hermitian operators in an $N$-level system,
we may always rewrite a product of spin operators as follows,
\begin{equation}
S^{\alpha}\hat{S}^{\beta}=\sum_{\gamma}c_{\gamma}\hat{T}^{\gamma}\label{eq:nonlinear_op_as_linear_combo}
\end{equation}
for some choice of $c_{\gamma}$. Thus
\[
\begin{alignedat}{1}\lim_{\lambda^{(N)}_{1}\rightarrow\infty}\left\langle \Psi\right|\hat{S}^{\alpha}\hat{S}^{\beta}\left|\Psi\right\rangle  & =\sum_{\gamma}c_{\gamma} \lim_{\lambda^{(N)}_1\rightarrow\infty}\left\langle \Psi\right|\hat{T}^{\gamma}\left|\Psi\right\rangle\\
 & =\sum_{\gamma}c_{\gamma}\left\langle \Psi\right|\hat{T}^{\gamma}\left|\Psi\right\rangle \\
 & =\sum_{\gamma}c_{\gamma}\Psi^{\gamma}.
\end{alignedat}
\]
Note that it was never necessary to invoke the factorization rule. This is in fact a
general feature of the large-$\lambda^N_1$ when considering an expectation value with respect
to a single SU$(N)$ coherent state: there is never any need to apply the factorization
rule or drop lower-order terms. The treatment of many-spin wave functions requires an
additional assumption, which is discussed in Section~\ref{sec:equations_of_motion}.

    As an illustration, we will consider the classical Hamiltonian that emerges under each limit for the following single-site, $S=1$ Hamiltonian,
    \begin{equation}
        \hat{\mathcal{H}} = D\hat{O}^{(2)}_{0} = D\left[3\left(\hat{S}^z\right)^2 - \hat{\mathbf{S}}^2\right],
    \end{equation}
    where $\hat{O}^{(2)}_{0}$ is the Stevens operator with a physical significance essentially identical to the Hamiltonian given in Eq.~\eqref{eq:hamiltonian_quad_aniso}, with the
    important distinction that it transforms as an irrep of the rotation group.
    We begin with the large-$S$ classical limit taken with respect to an arbitrary coherent states of SU$(2)$, $\left\vert\Omega\right\rangle$. The classical Hamiltonian
    will simply be the expectation value of the quantum Hamiltonian in this limit:
    \begin{equation}
        \mathcal{H}_{\rm cl} = \lim_{S\rightarrow\infty} \langle\Omega\vert\hat{\mathcal{H}}\vert\Omega\rangle
    \end{equation}
    To evaluate the first term of the Hamiltonian, we invoke the factorization rule Eq.~\eqref{eq:factorization_rule}:
    \begin{equation}
        \lim_{S\rightarrow\infty}\langle\Omega\vert\ \left(\hat{S}^z\right)^2 \vert\Omega\rangle = \langle\Omega\vert \hat{S}^z \vert\Omega\rangle^2.
    \end{equation}
    To calculate the expectation value, we insert identities into the expression as follows 
    \begin{equation}
        \left\langle\Omega\right\vert \hat{S}^z \left\vert\Omega\right\rangle = 
        \left\langle\Omega\right\vert U^\dagger_g U_g \hat{S}^z U_g^\dagger U_g \left\vert\Omega\right\rangle
    \end{equation}
    where we select $g$ to be the group action that transforms an arbitrary $\vert\Omega\rangle$ into the fully polarized state. Recognizing that the spin operators transform as a vector and that $\left\langle S \right\vert \hat{S}^\alpha \left\vert\ S \right\rangle = 0$ for $\alpha=x,y$, this reduces to
    \begin{equation}
        \left\langle S \right\vert R_g^{\mu\nu} \hat{S}^\nu \left\vert S \right\rangle =
        \left\langle S \right\vert R_g^{zz} \hat{S}^z \left\vert S \right\rangle = S\cos\theta
    \end{equation}
    
    The second term of the Hamiltonian is constant for all coherent states,
    \begin{equation}
        \lim_{S\rightarrow\infty} \langle\Omega\vert \hat{\mathbf{S}}^2\vert\Omega\rangle = \lim_{S\rightarrow\infty}S\left(S+1\right) = S^2,
    \end{equation}
    where for the last equality we have have dropped the term of linear order in $S$. 
    The classical Hamiltonian that emerges in the large-$S$ limit is therefore
    \begin{equation}
        \mathcal{H}_{\rm cl} = \lim_{S\rightarrow\infty}
        \langle \Omega \vert \hat{\mathcal{H}} \vert \Omega \rangle (\theta, \phi) = DS^2 \left( 3 \cos^2 \theta  - 1 \right).
        \label{eq:large_S_ham_example}
    \end{equation}
    This is the standard classical Hamiltonian that would be used for Landau-Lifshitz dynamics.
    It is also the appropriate starting for point determining the classical ground state
    for a traditional spin wave calculation, and the quantization of the small oscillations of this Hamiltonian would result in the standard spin wave theory.

    In the large-$\lambda^{(N)}_1$ limit, instead of considering SU$(2)$ coherent states, as given explicitly in Eq.~\eqref{eq:cs_parameterization}, we consider
    coherent states of SU$(N)$ where $N=2S+1$. For the example Hamiltonian, this corresponds to coherent states of $SU(3)$, i.e., the set of all pure states of a 3-level quantum
    system. These may be parameterized explicitly in terms of four angles: $\vert\Psi\left(\theta,\phi,\psi_1,\psi_2\right)\rangle$. The corresponding classical Hamiltonian is
    \begin{equation}
        \mathcal{H}_{\textrm{SU}(3)} = \lim_{\lambda^3_1\rightarrow\infty} \langle\Psi\vert \hat{\mathcal{H}} \vert\Psi\rangle (\theta,\phi,\psi_1,\psi_2) = \langle\Psi\vert \hat{\mathcal{H}} \vert\Psi\rangle (\theta,\phi,\psi_1,\psi_2)
    \end{equation}
    where we have taken advantage of the fact that the expectation value in the infinite representation limit of SU$(3)$ is identical to the expectation value in the fundamental representation -- there is no need to invoke the factorization rule or drop terms of lower order.
    We omit the explicit parameterization here; instead we refer to \cite{Hao21} for a more completely worked example and direct readers to \cite{nemoto2000generalized} for for a general parameterization scheme for SU($N$) coherent states.
    The key observation is that the Hamiltonian is now simply a classical function in four  real variables (angles) and it is exactly equal to the quantum expectation value with respect to the coherent state specified by those four angles. 
    
    As noted above, the coherent states of SU$(2)$ always exist as a submanifold of the coherent states of SU$(N)$. In particular, we may choose a parameterization such that 
    \begin{equation}
        \vert \Omega \rangle \left(\theta,\phi\right) = \vert \Psi \rangle \left(\theta,\phi,0,0\right).
    \end{equation}
    It is instructive to evaluate the classical Hamiltonian, as derived in using the large-$\lambda^{(3)}_1$ limit, while restricting the set of possible states to submanifold of coherent states of SU$(2)$.
    Using Eq.~\eqref{eq:cs_parameterization}, we find that an explicit parameterization
    of an SU$(2)$ coherent state for a 3-level system (spin-1 representation) is given
    by:
    \begin{equation}
        \vert\Omega\rangle \left(\theta,\phi\right) = \cos^2\frac{\theta}{2} e^{i\phi/2} |1\rangle + \sqrt{2}\cos\frac{\theta}{2}\sin\frac{\theta}{2}\vert0\rangle + \sin^2\frac{\theta}{2}e^{i\phi/2}\vert\ -1\rangle.
    \end{equation}
    This can then be used to directly evaluate the expectation value of $\hat{\mathcal{H}}$ in the fundamental representation of SU$(3)$. A simple calculation then shows,
    \begin{equation}
    \mathcal{H}_{\mathrm{SU}(3)}^\prime \left(\theta,\phi\right) = \lim_{\lambda^{(3)}_1\rightarrow\infty} \left\langle\Psi\right\vert \hat{\mathcal{H}} \left\vert\Psi\right\rangle\left(\theta,\phi,0,0\right) = \left\langle \Omega \right\vert \hat{\mathcal{H}} \left\vert \Omega \right\rangle \left(\theta,\phi\right) = \frac{3}{2}\cos^2 \theta - \frac{1}{2}.  
    \label{eq:large_lambda_ham_example}
    \end{equation}
    where the prime indicates that the Hamiltonian has been restricted to coherent states of SU$(2)$. Note that, for $S=1$, Eqs.~\eqref{eq:large_S_ham_example} and
    ~\eqref{eq:large_lambda_ham_example} differ by an overall factor of $\frac{1}{2}$.
    In Section \ref{sec:renorm_on-site_anisotropies} we will show that the proportionality factor between the two limits, for any $S$,
    will be $1-(2S)^{-1}$. 
    
    It is important to recognize that this $S^{-1}$ correction, and those
    introduced subsequently, have nothing to do with
    the introduction of quantum corrections. The control parameter used to  
    derive $\mathcal{H}_{\mathrm{SU}(3)}^\prime$ was $\lambda^{(3)}_1$ and quantum 
    corrections 
    would be introduced by terms in $(\lambda^{(3)}_1)^{-1}$. Here, $S^{-1}$ simply appears in the proportionality factor between two different classical limits.
    
    We also emphasize that the large-$\lambda^{(N)}_1$ Hamiltonian, when restricted to coherent
    states of SU$(2)$, is a classical Hamiltonian on the exact same phase space
    as the large-$S$ classical Hamiltonian. In other words, by simply renormalizing
    the large-$S$ Hamiltonian, we can recover the large-$\lambda^{(N)}_1$ Hamiltonian
    restricted to SU$(2)$ coherent states. 
    This is important because large-$\lambda^{(N)}_1$ Hamiltonian exactly reproduces
    the quantum expectation. It is in this sense
    that the renormalized classical Hamiltonian is more accurate than
    the traditional one whenever the Hamiltonian contains terms that are nonlinear in terms of the spin operators. Moreover, the procedure is straightforward and appropriate whenever whenever such a Hamiltonian retains a predominantly dipolar character, i.e., whenever the reduction of the dipole moment is small.

In what follows, we will derive general expressions for the classical equations of motion in both the large-$S$ and 
large-$\lambda^{(N)}_1$ limits. We will then
constrain the large-$\lambda^{(N)}_1$ classical limit to dipoles, i.e., to the coherent states
generated by linear combinations of spin operators only. The constrained
equations will then be compared to the
the large-$S$ limit.
We will find that the two results
are identical except that certain terms in the Hamiltonian have been renormalized. This
renormalization may be expressed in terms of powers of $S^{-1}$.
Indeed, these corrections capture precisely the terms
that are dropped when applying the factorization rule in the large-$S$ limit, Eq.~\eqref{eq:expectation_of_prod_with_corrections}. 

\subsection{Restricted equations of motion \label{sec:equations_of_motion}}

Let $\hat{\mathcal{H}}({\bf S})$ be a many-spin Hamiltonian
where ${\bf S}=(\hat{S}^x_{1},\hat{S}^y_{1},\hat{S}^z_{1},\ldots,\hat{S}^x_{n},\hat{S}^y_{n},\hat{S}^z_{n})$ and $n$ is the number of sites. 
We assume that the Hamiltonian may be written strictly as a polynomial in the spin operators on each site.
The associated classical equations of motion 
may be derived using the substitution rules outlined above with
the additional assumption that we neglect entanglement between sites.
This condition is enforced by restricting to a tensor
product basis of coherent states of SU$(2)$,
\[
\left|\Omega\right\rangle =\bigotimes_{j}\left|\Omega_{j}\right\rangle 
\]
where $j$ is the site index. 
This choice in turn restricts the time-evolution operator to be an element of 
tensor product $\bigotimes_{j}$SU(2)$_j$, where SU(2)$_j$ is the group of (local) rotations of the spin $j$.
Heisenberg's equations of motion may then be applied to each spin operator:
\begin{equation}
i\hbar\frac{d\hat{S}_{j}^{\alpha}}{dt} =\left[\hat{S}_{j}^{\alpha},\hat{\cal{H}}\left({\bf S}\right)\right]
=i\hbar\epsilon_{\alpha\beta\gamma}\frac{\partial\mathcal{\hat{H}}\left({\bf S}\right)}{\partial\hat{S}_{j}^{\beta}}\hat{S}_{j}^{\gamma}
\end{equation}
The operator derivative on the right is to be interpreted in the sense described in Appendix A, and involves an implicit contraction of the matrix indices for $\hat{S}_{j}^{\beta}$ and $\hat{S}_{j}^{\gamma}$.
The anti-symmetric Levi-Civita tensor arises from the commutation relations $\left[S^{\alpha},S^{\beta}\right]=i\hbar\epsilon_{\alpha\beta\gamma}S^{\gamma}$.
Formally, the commutation relations define the \emph{structure constants}
associated with basis for a Lie algebra, and $\epsilon_{\alpha\beta\gamma}$
are the structure constants of $\mathfrak{su}\left(2\right)$ with
the spin operators serving as a basis. 

If we define the classical Hamiltonian
\begin{equation}
\mathcal{H}_{\rm cl} \equiv \lim_{S\rightarrow\infty} \left\langle \Omega \right|\hat{\mathcal{H}}\left({\bf S}\right)\left|\Omega\right\rangle \label{eq:SU(2)-classical-hamiltonian}
\end{equation}
and apply the substitutions rule given in Eq. (\ref{eq:su2_basic_sub_rule}),
we may write,
\begin{equation}
\frac{d\Omega_{j}^{\alpha}}{dt}=\epsilon_{\alpha\beta\gamma}\frac{\partial\cal{H}_{\rm cl}}{\partial\Omega_{j}^{\beta}}\Omega_{j}^{\gamma},\label{eq:landau-lifshitz}
\end{equation}
which are precisely the LL equations. Note that calculating $\cal{H}_{\rm cl}$
demands attention to the details outlined in Section~\ref{sec:classical_limit}. In particular,
it may be necessary to apply the factorization rule and
drop terms of lower order in $S$ from groups of terms modified by a common coefficient. 
Detailed examples may be found in \cite{Hao21, Dahlbom22, Dahlbom22b}.

The generalized equations are derived in a directly analogous manner, but,
instead of using coherent states of SU$(2)$, the limit is taken with respect to coherent states of
SU$(N)$, where typically $N=2S+1$ and $S$ is the total effective spin
on each site. The Hilbert space is first restricted
to a tensor product basis of SU$(N)$ coherent states,
\[
\left|\Psi\right\rangle =\bigotimes_{j}\left|\Psi_{j}\right\rangle.
\]
We consider the same Hamiltonian as before.
Note, however, that all terms of the Hamiltonian that are nonlinear
in the spin operators for a single site may now be expressed as linear
combinations of the generators of SU$(N)$, as in Eq.~\eqref{eq:nonlinear_op_as_linear_combo}.
To express this fact notationally, we here call the Hamiltonian $\hat{H}\left({\bf T}\right)$,
where ${\bf T}$ is the vector of all generators of SU$(N)$
on all sites. 

Heisenberg's equations of motion are again applied, now to each generator
$T_{j}^{\alpha}$,
\begin{equation}
i\hbar\frac{d\hat{T}_{j}^{\alpha}}{dt}=\left[\hat{T}_{j}^{\alpha},\hat{\mathcal{H}}\left({\bf T}\right)\right]
 =i\hbar f_{\alpha\beta\gamma}\frac{\partial\mathcal{\hat{H}}\left({\bf T}\right)}{\partial\hat{T}_{j}^{\beta}}\hat{T}_{j}^{\gamma},
\label{eq:heisenberg_eq_in_T}
\end{equation}
with an implicit contraction on the matrix elements of $\hat{T}_{j}^{\beta}$ and $\hat{T}_{j}^{\gamma}$. Now the structure constants $f_{\alpha\beta\gamma}$ of $\mathfrak{su}\left(N\right)$ appear,
arising from the commutator $\left[\hat{T}^{\alpha},\hat{T}^{\beta}\right]=i\hbar f_{\alpha\beta\gamma}\hat{T}^{\gamma}$.
The classical Hamiltonian is next defined using the large-$\lambda^N_1$ limit. Since 
the Hamiltonian is linear in the generators, we may apply Eq.~\eqref{eq:sun_sub_rule} and simply write,
\begin{equation}
\mathcal{H}_{\mathrm{SU}\left(N\right)}\equiv\left\langle \Psi\right|\mathcal{\mathcal{H}}\left({\bf T}\right)\left|\Psi\right\rangle. \label{eq:SU(N)-classical-hamiltonian}
\end{equation}
Using this definition and applying our substitution rules to Eq.~\eqref{eq:heisenberg_eq_in_T}, we find,
\begin{equation}
\frac{d\Psi_{j}^{\alpha}}{dt}=f_{\alpha\beta\gamma}\frac{\partial\mathcal{H}_{SU\left(N\right)}}{\partial\Psi_{j}^{\beta}}\Psi_{j}^{\gamma}.\label{eq:generalized-landau-lifshitz}
\end{equation}
These generalized equations have the same form as the traditional
LL equations of Eq.~\eqref{eq:landau-lifshitz}. However, the number of
dynamical variables on each site has increased to $N^{2}-1$, and the
cross product, which comes from the structure constants of $\mathfrak{su}(2)$,
has been generalized using the structure constants of $\mathfrak{su}(N)$.
Additionally, since the new classical Hamiltonian
is defined by taking an expectation with respect to the tensor product of SU$(N)$ coherent states, it is never necessary to invoke the factorization rule or drop lower-order terms. 

While the generalized LL dynamics of Eq.~\eqref{eq:generalized-landau-lifshitz}
has many advantages, it may also introduce sharp modes that would
be washed out in multi-magnon continua when higher-order corrections in $\lambda^N_1$
are included. We therefore propose restricting the generalized dynamics
to a dynamics of dipoles only. The Appendix B of this SM
contains a direct calculation showing that the restricted
dynamics takes the form,
\begin{equation}
\frac{d\Omega_j^{\alpha}}{dt}=\epsilon_{\alpha\beta\gamma}\frac{\partial\mathcal{H}_{\textrm{SU}\left(N\right)}}{\partial\Omega_j^{\beta}}\Omega_j^{\gamma}.\label{eq:restricted_dynamics_final}
\end{equation}
These equations are identical to the LL equations except that
$\mathcal{H}_{\rm cl}$ has been replaced with ${\cal H}_{\textrm{SU}(N)}$, i.e.,
the classical Hamiltonian has been calculated using SU$(N)$ coherent states 
in the large-$\lambda^N_1$ limit. Since the manifold of
SU$(N)$ coherent states contains within it the submanifold
of SU$(2)$ coherent states generated by the spin operators,
restricting $\mathcal{H}_{\rm{SU}\text{\ensuremath{\left(N\right)}}}$
to this submanifold is a natural operation. To make practical use
of Eq.~\eqref{eq:restricted_dynamics_final}, it is necessary to evaluate
$\mathcal{H}_{\textrm{SU}(N)}$ on this restricted space and see
how it differs from $\mathcal{H}_{cl}$. Recall that,
for any term that is linear in the spin operators, the large-$S$
and large-$\lambda^N_1$ classical limits agree. The remainder
of the SM is therefore dedicated to examining the difference between
these limits when they are applied to a large class of nonlinear terms.

\subsection{Renormalization of on-site anisotropies \label{sec:renorm_on-site_anisotropies}}

On-site anisotropies are a common source of terms that are nonlinear in the spin operators.
A useful basis for expressing such anisotropies
is the Stevens operators, $\hat{O}^{(k)}_{q}$. Like the spherical tensors, the Stevens operators (for each fixed $k$) are an irrep of SO(3). That is, under a physical 3D rotation, each $\hat{O}^{(k)}_{q}$ transforms to linear combination of $\hat{O}^{(k)}_{q'}$ involving $q' = -k, \dots k$. There is a simple linear relationship between Stevens operators and spherical tensors~\cite{rudowicz1985transformation}. Stevens operators, unlike spherical tensors, are Hermitian. The set of all $\hat{O}^{(k)}_{q}$ for $k=1,\dots N-1$ and $q = -k, \dots k$ can be used as a basis for the $N^2-1$ generators of SU($N$).

Following the approach described in Section~\ref{sec:classical_limit}, the classical limit of a Stevens
operator may be found by evaluating its expectation value with respect to a coherent
state in the infinite representation. The only states that are relevant for the restricted dynamics, Eq.~\eqref{eq:restricted_dynamics_final},
are the coherent states of SU$(2)$ generated by the spin operators.
We label these $\vert\Omega\rangle$. Our task, then, is to evaluate
\[
\left\langle \Omega \right\vert \hat{O}^{(k)}_{q} \left\vert \Omega \right\rangle
\]
in both the large-$\lambda^{(N)}_1$ and large-$S$ limits.

By construction, the limiting procedures will produce results that transform according
to identical irreps and will be defined on the same domain, namely the two-sphere. They  
will therefore be proportional. Moreover, the proportionality constant will be independent
of $q$ \footnote{To show this formally,
one may first consider spherical tensors and
spherical harmonics. The expectation values of the spherical tensors
with respect to coherent states will transform according to the same
irreps as the spherical harmonics and will therefore be proportional.
Repeated application of the orthogonality theorem for irreps will
show proportionality that is independent of $q$. The result is analogous
to the Wigner-Eckart theorem. The argument carries over to Stevens
operators because they are related to the spherical tensors by a linear
transformation that does not mix different $k$s. The transformation is given explicitly in
\cite{rudowicz1985transformation}}. It follows that
\begin{equation}
    \left\langle\Omega\right\vert \hat{O}^{(k)}_{q} \left\vert\Omega\right\rangle = c_k \left[\lim_{S\rightarrow\infty}\left\langle\Omega\right\vert \hat{O}^{(k)}_{q} \left\vert\Omega\right\rangle\right],\label{eq:ck_equation}
\end{equation}
for any $|\Omega\rangle$, where the expectation on the left is precisely what emerges in the large-$\lambda^{(N)}_1$ limit, Eq.~\eqref{eq:sun_sub_rule}.
To determine $c_{k}$, we may simply evaluate both sides of this equation using
any SU$(2)$ coherent state. The $z$-polarized state, $|S\rangle$, is a particularly
convenient choice.

We will restrict our attention to Stevens operators
$\hat{O}^{(k)}_{q}$ where $k$ is even (to respect time-reversal symmetry)
and less than or equal to 6. Since the proportionality constant is
independent of $q$, we will take $q=0$. The relevant Stevens operators
are then:

\[
\begin{alignedat}{1}
\hat{O}^{(2)}_{0} =\ & 3\left(\hat{S}^{z}\right)^{2}-\mathbf{S}^{2}\\
\hat{O}^{(4)}_{0} =\ & 35\left(\hat{S}^{z}\right)^{4}-\left(30\mathbf{S}^{2}-25\right)\left(\hat{S}^{z}\right)^{2}+3\mathbf{S}^{4}-6\mathbf{S}^{2}\\
\hat{O}^{(6)}_{0} =\ & 231\left(\hat{S}^{z}\right)^{6}-\left(315\mathbf{S}^{2}-735\right)\left(\hat{S}^{z}\right)^{4} \\ & \quad+ \left(105\mathbf{S}^{4}-525\mathbf{S}^{2}+294\right)\left(\hat{S}^{z}\right)^{2} \\ & \quad-5\mathbf{S}^{6}+40\mathbf{S}^{4}-60\mathbf{S}^{2}
\end{alignedat}
\]

Beginning with $\hat{O}^{(2)}_{0}$, the left-hand side of Eq.~\eqref{eq:ck_equation} is evaluated directly,
\[
\begin{alignedat}{1}
\left\langle S\right\vert \hat{O}^{(2)}_{0} \left\vert S\right\rangle&=3\langle S\vert\left(\hat{S}^{z}\right)^{2}\vert S\rangle -\left\langle S\left|\mathbf{S}^{2}\right|S\right\rangle\\ &=3S^{2}-S(S+1).
\end{alignedat}
\]
The large-$S$ limit is calculated as follows,
\[
\begin{alignedat}{1}
\lim_{S\rightarrow\infty}\left\langle S \right\vert \hat{O}^{(2)}_{0} \left\vert S\right\rangle&=3\left\langle S\left|\hat{S}^{z}\right|S\right\rangle ^{2}\\&\quad\quad-\left(\left\langle S\left|\hat{S}^{x}\right|S\right\rangle ^{2}+\left\langle S\left|\hat{S}^{y}\right|S\right\rangle ^{2}+\left\langle S\left|\hat{S}^{z}\right|S\right\rangle ^{2}\right)\\&=2S^{2},
\end{alignedat}
\]
where we have made use of the factorization rule. We can now solve for $c_{2}$:
\begin{equation}
    3S^{2}-S(S+1)=c_{2}\left(2S^{2}\right)\implies c_{2}=1-\frac{1}{2S}. \label{eq:c2_derivation}
\end{equation}
The same procedure can be applied to find $c_{4}$ and $c_{6}$. Evaluating
expectation values, we find,
\[
\begin{alignedat}{1}
 \left\langle S\right\vert \hat{O}^{(4)}_{0} \left\vert S\right\rangle& =8S^{4}-24S^{3}+22S^{2}-6S\\
 \left\langle S\right\vert \hat{O}^{(6)}_{0} \left\vert S\right\rangle& =16S^{6}-120S^{5}+340S^{4}-550S^{3}+274S^{2}-60S,
\end{alignedat}
\]
Taking care to drop terms of lower order in $S$, the large-$S$ limit yields
\[
\begin{alignedat}{1}
\lim_{S\rightarrow\infty} \left\langle S \right\vert \hat{O}^{(4)}_{0} \left\vert S\right\rangle& =8S^{4}\\
\lim_{S\rightarrow\infty} \left\langle S \right\vert \hat{O}^{(6)}_{0} \left\vert S\right\rangle& =16S^{6}.
\end{alignedat}
\]
It is now straightforward to solve for the remaining $c_{k}$:
\[
\begin{alignedat}{1}
c_{2} & =1-\frac{1}{2}S^{-1}\\
c_{4} & =1-3S^{-1}+\frac{11}{4}S^{-2}-\frac{3}{4}S^{-3}\\
c_{6} & =1-\frac{15}{2}S^{-1}+\frac{85}{4}S^{-2}-\frac{225}{8}S^{-3}+\frac{137}{8}S^{-4}-\frac{15}{4}S^{-5}
\end{alignedat}
\]

\subsection{Renormalization of Biquadratic interactions\label{sec:renorm_biquadratic}}

$n$-spin interactions with $n>2$ are another common source of nonlinear terms. If we restrict
attention to interactions that respect time-reversal symmetry, requiring $n$ to be even, the simplest
example is the isotropic biquadratic interaction:
\[
\left({\bf S}_{1}\cdot{\bf S}_{2}\right)^{2}
\]
We will directly calculate this term in both classical limits
to determine the appropriate renormalization. This example is
provided as a template for more general $n$-spin interactions.

When the biquadratic interaction is fully expanded, all resulting terms will have
degree 4. The large-$S$ classical limit may therefore be calculated
simply by replacing each spin operator with its expectation value taken with
respect to a coherent state on each site: $S_{j}^{\alpha}\rightarrow\left\langle \Omega_{j}\right|S_{j}^{\alpha}\left|\Omega_{j}\right\rangle =\Omega_{j}^{\alpha}$.
The result is then
\begin{equation}
\left({\bf \Omega}_{1}\cdot{\bf \Omega}_{2}\right)^{2}.\label{eq:biquad_su(2)_cl}
\end{equation}

The large-$\lambda^N_1$ classical limit may be calculated by directly taking
the expectation value of $\left({\bf S}_{1}\cdot{\bf S}_{2}\right)^{2}$
with respect to the tensor product of two SU$(2)$ coherent states, which
will be denoted $\left\vert\Omega_1 \Omega_2\right\rangle$. We begin by exploiting the invariance of the biquadratic interaction under SU$\left(2\right)$ rotations,
\[
\begin{split}
& \left\langle \Omega_{1},\Omega_{2}\right|\left({\bf S}_{1}\cdot{\bf S}_{2}\right)^{2}\left|\Omega_{1}\Omega_{2}\right\rangle\\
& \quad\quad =\left\langle \Omega_{1},\Omega_{2}\right|U_{g}^{\dagger}U_{g}\left({\bf S}_{1}\cdot{\bf S}_{2}\right)^{2}U_{g}^{\dagger}U_{g}\left|\Omega_{1},\Omega_{2}\right\rangle\\
& \quad\quad=\left\langle \Omega_{1},\Omega_{2}\right|U_{g}^{\dagger}\left({\bf S}_{1}\cdot{\bf S}_{2}\right)^2 U_{g}\left|\Omega_{1},\Omega_{2}\right\rangle 
\end{split}
\]
where $U_{g}=e^{i{\bf n\cdot\left({\bf S}_{1}+{\bf S}_{2}\right)}\alpha}$
is a global spin rotation. We may specifically choose $U_{g}$ to
transform $\Omega_{1}$ to be the maximal weight state $\left|S\right\rangle $:
\[
U_{g}\left|\Omega_{1},\Omega_{2}\right\rangle =\left|S,\Omega_{2}^{\prime}\right\rangle 
\]
We additionally have the freedom to specify that this transformation
place the vector ${\bf \Omega}_{2}^{\prime}=\left\langle \Omega_{2}\right|{\bf S}_{2}\left|\Omega_{2}\right\rangle $
in the plane orthogonal to the $y$-axis. ${\bf \Omega}_{2}^{\prime}$
will then have the general form $\left(S\sin\theta,0,S\cos\theta\right)$,
where $\theta$ is the angle between ${\bf \Omega}_{1}^{\prime}$
and ${\bf \Omega}_{2}^{\prime}$. In particular, we have ${\bf \Omega}_{1}\cdot{\bf \Omega}_{2}={\bf \Omega}_{1}^{\prime}\cdot{\bf \Omega}_{2}^{\prime}=S^{2}\cos\theta$.
We may then write
\[
\left\langle \Omega_{1},\Omega_{2}\right|\left({\bf S}_{1}\cdot{\bf S}_{2}\right)^{2}\left|\Omega_{1}\Omega_{2}\right\rangle =\left\langle S,\Omega_{2}^{\prime}\right|\left({\bf S}_{1}\cdot{\bf S}_{2}\right)^{2}\left|S,\Omega_{2}^{\prime}\right\rangle .
\]

We introduce another unitary transformation that changes the relative
angle between ${\bf \Omega}_{1}^{\prime}$ and ${\bf \Omega}_{2}^{\prime}$.
This can be achieved by rotating the second spin about the $y$-axis.
Specifically, we will select a transformation which carries ${\bf \Omega}_{2}^{\prime}$
into the maximal weight state as well: $U_{r}=e^{-iS_{2}^{y}\theta}$. Then,
\[
\begin{alignedat}{1}\left\langle \Omega_{1},\Omega_{2}\right|\left({\bf S}_{1}\cdot{\bf S}_{2}\right)^{2}\left|\Omega_{1}\Omega_{2}\right\rangle  & =\left\langle S,\Omega_{2}^{\prime}\right|\left({\bf S}_{1}\cdot{\bf S}_{2}\right)^{2}\left|S,\Omega_{2}^{\prime}\right\rangle \\
 & =\left\langle S,S\right|U_{r}\left({\bf S}_{1}\cdot{\bf S}_{2}\right)^{2}U_{r}^{\dagger}\left|S,S\right\rangle \\
 & =\left\langle S,S\right|\left({\bf S}_{1}\cdot U_{r}{\bf S}_{2}U_{r}^{\dagger}\right)^{2}\left|S,S\right\rangle.
\end{alignedat}
\]
Since ${\bf S}_{2}$ is a vector operator, its transformation under
$U_{r}$  is equivalent to a transformation on the left by an orthogonal
matrix $R_{r}$,
\[
U_{r}{\bf S}_{2}U_{r}^{\dagger}=R_{r}^{\mu\nu}\hat{S}^{\nu}.
\]
Inserting this into our calculation, we find
\[
\begin{alignedat}{1}
& \left\langle S,S\right|\left({\bf S}_{1}\cdot U_{r}{\bf S}_{2}U_{r}^{\dagger}\right)^{2}\left|S,S\right\rangle\\
& \quad=\left\langle S,S\right|\left(\hat{S}_{1}^{\mu}R_{r}^{\mu\nu}\hat{S}_{2}^{\nu}\right)^{2}\left|S,S\right\rangle \\
 & \quad=S^{4}(R_{r}^{zz})^{2}+\frac{S^{2}}{4}(1-R_{r}^{xx})+\frac{S^{2}}{4}[(R_{r}^{xx})^{2}-R_{r}^{xx}]+(R^{xz})^{2}S^{3}\\
 & \quad=S^{4}(\cos{\theta})^{2}+\frac{S^{2}}{4}(1-\cos{\theta})+\frac{S^{2}}{4}\cos{\theta}(\cos{\theta}-1)\\
 &\quad\quad +(1-\cos^{2}{\theta})S^{3}\\
 & \quad=\left(S^{4}-S^{3}+\frac{S^{2}}{4}\right)(\cos{\theta})^{2}-\frac{S^{2}}{2}\cos{\theta}+S^{3}+\frac{S^{2}}{4}\\
 & \quad=\left(1-\frac{1}{S}+\frac{1}{4S^{2}}\right)({\bf \Omega}_{1}\cdot{\bf \Omega}_{2})^{2}-\frac{1}{2}{\bf \Omega}_{1}\cdot{\bf \Omega}_{2}+S^{3}+\frac{S^{2}}{4}.
\end{alignedat}
\]
By comparing with Eq.~\eqref{eq:biquad_su(2)_cl}, it is clear that
not only must $({\bf \Omega}_{1}\cdot{\bf \Omega}_{2})^{2}$ be renormalized---by a rather substantial amount for realistic spin values---but it is also necessary to introduce a quadratic correction, the order of which is similar to the quartic term. 

\subsection{Summary of RCS Theory}

We have examined the procedure for deriving the a classical Hamiltonian under two different classical limits, the traditional large-$S$ limit and the large-$\lambda_1^{(N)}$ limit. The RCS Hamiltonian , $\tilde{\mathcal{H}}_{\rm cl} $, is obtained by constraining the latter limit to the same phase space as the large-$S$ Hamiltonian, $\mathcal{H}_{\rm cl}$. $\tilde{\mathcal{H}}_{\rm cl} $ and $\mathcal{H}_{\rm cl}$
are identical except that terms that are nonlinear in the spin operators have been renormalized in $\tilde{\mathcal{H}}_{\rm cl} $
(and, in the case of $n$-spin interactions, there may appear additional terms). This is significant, because the RCS Hamiltonian is in a basic sense more accurate, as it yields the exact expectation values for quantum operators for any given SU(2) coherent state (i.e., dipolar state). 

It is important to stress that $\mathcal{H}_{\rm cl}$ and $\tilde{\mathcal{H}}_{\rm cl} $ are classical Hamiltonians
defined on the same phase space, namely the manifold of SU$(2)$ coherent states. Thus, any operation one might perform with 
$\mathcal{H}_{\rm cl} $
can be performed just as well with 
the RCS Hamiltonian
$\tilde{\mathcal{H}}_{\rm cl} $. For example, consider the following spin-$S$ quantum Hamiltonian,

\begin{equation}
    \hat{{\cal H}} = J {\bm S}_1 \cdot {\bm S}_2 + D \left[ \left(\hat{S}^z_1\right)^2 + \left(\hat{S}^z_2\right)^2\right],
\end{equation}
where we assume that $\vert J \vert \gg \vert D \vert$ so that the dipole possess significant stiffness. (If this were not the case, the full SU($N$) treatment \cite{Hao21}, which better models quadrupolar excitations, would be the appropriate approach.) The classical, large-$S$ Hamiltonian is simply
\begin{equation}
    \mathcal{H}_{\rm cl} = J{\bf \Omega}_1\cdot {\bf \Omega}_2 + D\left[ \left(\Omega^z_1\right)^2 + \left(\Omega^z_2\right)^2 \right],
\end{equation}
and the RCS Hamiltonian is
\begin{equation}
    \tilde{\mathcal{H}}_{\rm cl} = J{\bf \Omega}_1\cdot {\bf \Omega}_2 + \tilde{D}\left[ \left(\Omega^z_1\right)^2 + \left(\Omega^z_2\right)^2 \right],
\end{equation}
where $\tilde{D} = (1 -1/2S)D$. The rescaling follows Eq.~\eqref{eq:c2_derivation} and the observation that $(S^z)^2 \propto O^{(2)}_{0}$ if constant contributions are disregarded.
Both Hamiltonians may now be used to, for example, to calculate partition functions:
\begin{equation}
\begin{alignedat}{2}
\mathcal{Z}_{\rm cl} &= \int d\Omega_1 D\Omega_2 e^{-\beta\mathcal{H}_{\rm cl}\left({\bf \Omega}_1\cdot{\bf \Omega}_2\right)} = \int d\Omega_1 d\Omega_2 e^{-\beta\left(J{\bf \Omega}_1\cdot{\bf \Omega}_2 + D\left[ \left(\Omega^z_1\right)^2 + \left(\Omega^z_2\right)^2 \right] \right)}\\
\tilde{\mathcal{Z}}_{\rm cl} &= \int d\Omega_1 D\Omega_2 e^{-\beta\tilde{\mathcal{H}}_{\rm cl}\left({\bf \Omega}_1\cdot{\bf \Omega}_2\right)} = \int d\Omega_1 d\Omega_2 e^{-\beta\left(J{\bf \Omega}_1\cdot{\bf \Omega}_2 + \tilde{D}\left[ \left(\Omega^z_1\right)^2 + \left(\Omega^z_2\right)^2 \right] \right)}.\\
\end{alignedat}
\end{equation}
Observe that the only dependence of $\mathcal{Z}_{\rm cl}$ on $S$ is through the magnitude of the classical vector $\vert {\bf \Omega}_j\vert = \left\langle{\bf \Omega}_j \vert {\bf S}_j \vert {\bf \Omega}_j \right\rangle$. In contrast, $\tilde{\mathcal{Z}}_{\rm cl}$ includes an extra $S$-dependence in the renormalized single-ion anisotropy. The relationship between these partition functions is simply
\begin{equation}
\mathcal{\tilde{Z}}_{\rm cl}\left(J, D\right) = \mathcal{Z}_{\rm cl}\left(J, \tilde{D}\right)
\end{equation}
More generally, for any spin Hamiltonian with parameters ${p_j}$, the relationship
between the resulting classical parition functions, RCS and large-$S$, will be
\begin{equation}
    \tilde{\mathcal{Z}}_{\rm cl}\left(p_1,\ldots,p_m\right) = \mathcal{Z}_{\rm cl}\left(\tilde{p}_1,\ldots,\tilde{p}_m\right).
\end{equation}
where ${\tilde{p}_j}$ are renormalized parameters.

Similarly, both $\mathcal{H}_{\rm cl}$ and $\tilde{\mathcal{H}}_{\rm cl}$ may be used as the classical Hamiltonian in the the Landau-Lifshitz equations, or as the starting point for a traditional spin wave calculation. In the former case, the  control parameter is $1/S$. In contrast,  the spin wave theory based on  $\tilde{\mathcal{H}}_{\rm cl}$  must be regarded as an approximation to the result that is obtained by using generalized SU$(N)$ spin wave theory (whose control parameter is $1/\lambda_1^{(N)}$) and integrating out the non-dipolar modes. 



\subsection*{Appendix A: Operator derivative}

Consider the SU($N$) Lie algebra with generators $\hat{T}^{\alpha}$ that
satisfy
\begin{equation}
[\hat{T}^{\alpha},\hat{T}^{\beta}]=if_{\alpha\beta\gamma}\hat{T}^{\gamma}.
\end{equation}
In this appendix, we will assume the $\hat{T}^\alpha$ are matrices in some
chosen basis.
Further, we can assume that the Hamiltonian $\hat{\mathcal{H}}$ is some arbitrary polynomial
of $\hat{T}^{\alpha}$. For the purposes of this study, we may particularly
assume the Hamiltonian is restricted to the spin operators, $\hat{T}^\alpha$
with $\alpha=1,2,3$, but the argument will be made in general.
The commutator, $[\hat{T}^{\alpha},\hat{\mathcal{H}}]$
is closely related to the derivative $d\hat{\mathcal{H}}/d\hat{T}^{\alpha}$, in a way that
we will now make explicit. For this argument, it suffices to consider
a single term in the polynomial, 
\begin{equation}
\hat{A}=\hat{T}^{\beta_{1}}\hat{T}^{\beta_{2}}\dots \hat{T}^{\beta_{n}}.
\end{equation}
Each index $\beta_{i}$ is selecting one of the $N^{2}-1$ generators
of SU($N$). 

The Lie bracket satisfies a generalization of the ``product rule'' familiar for derivatives,
\begin{equation}
[\hat{T}^{\alpha}, \hat{A}]=\sum_{i=1}^{n}(\hat{T}^{\beta_{1}}\dots \hat{T}^{\beta_{i-1}})[\hat{T}^{\alpha},\hat{T}^{\beta_{i}}](\hat{T}^{\beta_{i+1}}\dots \hat{T}^{\beta_{n}}).
\end{equation}
Using the SU($N$) structure constants, this becomes
\begin{equation}
[\hat{T}^{\alpha},\hat{A}]=if_{\alpha\beta\gamma}\sum_{i=1}^{n}(\hat{T}^{\beta_{1}}\dots \hat{T}^{\beta_{i-1}})(\delta_{\beta,\beta_{i}}\hat{T}^{\gamma})(\hat{T}^{\beta_{i+1}}\dots \hat{T}^{\beta_{n}}),\label{eq:commute0}
\end{equation}
with $\delta_{\beta,\beta_{i}}$ the usual Kronecker delta. Although
the generators do not commute, we can move the $\hat{T}^{\gamma}$ operator
to the right by making the matrix elements explicit,
\begin{equation}
[\hat{T}^{\alpha},\hat{A}]_{kl}=if_{\alpha\beta\gamma}\left[\sum_{i=1}^{n}(\hat{T}^{\beta_{1}}\dots \hat{T}^{\beta_{i-1}})_{km}\delta_{\beta,\beta_{i}}(\hat{T}^{\beta_{i+1}}\dots \hat{T}^{\beta_{n}})_{nl}\right]\hat{T}_{mn}^{\gamma},\label{eq:commut}
\end{equation}
with summation over repeated matrix indices $m,n$ implied.

Now we will consider ordinary partial derivatives with respect to
matrix elements of the generators, e.g.,
\begin{equation}
\partial \hat{T}_{kl}^{\beta_{i}}/\partial \hat{T}_{mn}^{\beta}=\delta_{\beta_{i}\beta}\delta_{km}\delta_{ln}.
\end{equation}
Using the product rule for derivatives, the term in square brackets
of Eq.~(\ref{eq:commut}) is recognized as the usual partial derivative
of the matrix $A$,
\begin{equation}
[\hat{T}^{\alpha}, \hat{A}]=if_{\alpha\beta\gamma}\frac{\partial \hat{A}}{\partial \hat{T}_{mn}^{\beta}} \hat{T}_{mn}^{\gamma}.
\end{equation}
By linearity, a similar result holds for the full Hamiltonian $H$,
which is a sum over terms of the form of $\hat{A}$.

In taking the classical limit, we will calculate the expectation of
both sides of Eq.~(\ref{eq:commute0}) with respect to a product
state. This allows us to ignore matrix ordering, e.g.,
\begin{equation}
\langle \hat{T}^{\beta_{1}}\hat{T}^{\beta_{2}}\dots\rangle=\langle \hat{T}^{\beta_{1}}\rangle\langle \hat{T}^{\beta_{2}}\rangle\dots=\Omega^{\beta_{1}}\Omega^{\beta_{2}}\dots.
\end{equation}
The result is,
\begin{align}
\langle[\hat{T}^{\alpha},\hat{\mathcal{H}}]\rangle & =if_{\alpha\beta\gamma}\sum_{i=1}^{n}(\Omega^{\beta_{1}}\dots\Omega^{\beta_{i-1}})(\delta_{\beta,\beta_{i}}\Omega^{\gamma})(\Omega^{\beta_{i+1}}\dots\Omega^{\beta_{n}})\nonumber \\
 & =if_{\alpha\beta\gamma}\frac{\partial\langle \hat{\mathcal{H}}\rangle}{\partial\Omega^{\beta}}\Omega^{\gamma},
\end{align}
The second equality is justified by the usual product rule for derivatives,
in direct analogy to the product rule for used in Eq.~(\ref{eq:commute0}).

\subsection*{Appendix B: Constrained generalized dynamics}

Consider a generic spin Hamiltonian $\hat{\mathcal{H}}$,
which, for simplicity, we will take to be a single-site Hamiltonian; the generalization 
to a many-spin problem proceeds in the same way described in Section~\ref{sec:classical_limit}.
We will be considering expectation values with respect to the SU$(2)$ coherent
states generated by the spin operators. As described in Section~\ref{sec:generalized_cs},
these SU$(2)$ coherent states will be considered as a submanifold of SU$(N)$ coherent states, so
we are free to express the Hamiltonian
as a linear combination of generators
$T^{\alpha}$ belonging to the Lie algebra $\mathfrak{su}\left(N\right)$,
\begin{equation}
\mathcal{\hat{H}}=\sum_{\alpha=1}^{N^{2}-1}b_{\alpha}\hat{T}^{\alpha},\label{eq:generic_SU(N)_hamiltonian}
\end{equation}
for real coefficients $b_{\alpha}$.

We now propose another dynamics
generated by a constrained Hamiltonian, $\hat{h}$, which we will define as a linear
combination of the spin operators only. Since we have chosen $\hat{T}^{\alpha}$
for $\alpha=1,2,3$ to be the spin operators Eq. (\ref{eq:T1-3_as_S1-3}), 
we may write this as:
\begin{equation}
\hat{h}=\sum_{\alpha=1}^{3}c_{\alpha}\hat{T}^{\alpha}\label{eq:generic_restricted_hamiltonian}.
\end{equation}
This operator will generate group actions in $SU$$\left(2\right)$,
as in Eq.~\eqref{eq:arbitrary_SU(2)_cs}, and represents
the generic form of any closed dynamics on the SU$(2)$ coherent states of interest.


Recall from Eq.~\eqref{eq:omega_vector} that the Bloch construction is a bijective map between SU$(2)$ coherent states and points on the sphere. We may
therefore consider the expectation value of an operator with respect to an SU$(2)$
coherent state as a function on the sphere. In particular, we define
\[
\mathcal{H}_{\textrm{SU}(N)}\left(\Omega^x, \Omega^y, \Omega^z\right) \equiv \left\langle \Omega \right\vert \hat{\mathcal{H}} \left\vert \Omega\right\rangle.
\]
 The ``SU$(N)$'' subscript has been added because this expression is precisely $\mathcal{H}_{\textrm{SU}(N)}$ as defined in Eq.~\eqref{eq:SU(N)-classical-hamiltonian} when restricted to the sphere.
 
We wish to examine the time evolution of $\mathcal{H}_{\textrm{SU}(N)}$, the expected value of our full Hamiltonian, under the dynamics generated by our restricted Hamiltonian, $\hat{h}$. We begin by applying the chain rule:
\begin{equation}
\frac{d \mathcal{H}_{\textrm{SU}(N)}}{dt} = \sum_{\alpha=1}^{3}\frac{\partial \mathcal{H} _{\textrm{SU}(N)}}{\partial \Omega^\alpha}
\frac{d \Omega^\alpha}{dt}.
\label{eq:time_evolution_chain_rule}
\end{equation}

Noting that $\Omega^\alpha = \left\langle \Omega \right \vert T^\alpha \left\vert \Omega \right\rangle$,
the time evolution of this expression under $\hat{h}$ can be found using Heisenberg's equations of motion for the operator $\hat{T}^\alpha$,
\begin{equation}
    \frac{d\Omega^\alpha}{dt}
    = \left\langle \Omega\left|\frac{d\hat{T}^{\alpha}}{dt} \right|\Omega\right\rangle
    =\frac{-i}{\hbar}\left\langle \Omega\left|\left[\hat{T}^{\alpha},\hat{h}\right]\right|\Omega\right\rangle.\\
\end{equation}
The commutator may be evaluated using the expansion in Eq.~\eqref{eq:generic_restricted_hamiltonian} and invoking the SU($N$) structure constants,
\begin{equation}
    \left[\hat{T}^{\alpha},\hat{h}\right] = \sum_{\beta=1}^{3} c_\beta \left[\hat{T}^{\alpha},\hat{T}^{\beta}\right] = i f_{\alpha \beta \gamma} c_\beta T^{\gamma}
\end{equation}
Substituting these two results into Eq.~\eqref{eq:time_evolution_chain_rule} yields a classical dynamics,
\begin{equation}
\frac{d\Omega^\alpha}{dt} = \sum_{\beta=1}^{3}\sum_{\gamma=1}^{N^{2}-1}f_{\alpha\beta\gamma}c_{\beta}\Omega^\gamma.
\end{equation}

Since we have chosen the first three $\hat{T}^{\alpha}$ to be spin operators,
it follows $f_{\alpha\beta\gamma}=\epsilon_{\alpha\beta\gamma}$ for
indices with values between $1$ and $3.$
Moreover, the spin operators in fact form a subalgebra in the set of generators, so the $\gamma$-sum
never need extend beyond three when both operators in the commutator have indices between one and three. Using this fact while substituting the last result into Eq.~(\ref{eq:time_evolution_chain_rule}), we find

\begin{equation}
\frac{d \mathcal{H}_{\textrm{SU}(N)}}{dt} = \sum_{\alpha, \beta, \gamma=1}^{3}\epsilon_{\alpha\beta\gamma}\frac{\partial\mathcal{H}_{\textrm{SU}(N)}}{\partial \Omega^\alpha}c_\beta\Omega^\gamma
\end{equation}
This has the form of a scalar triple product. To ensure
that the dynamics generated by $\hat{h}$ keeps the expected value
of the original Hamiltonian constant, we wish to determine the values
of $c_{\beta}$ such that the right-hand side is always zero. From
geometric consideration, this means the $c_{\beta}$ must be coplanar
with the other two vectors:
\[
c_\alpha = \kappa\left(t\right)\frac{\partial\mathcal{H}_{\textrm{SU}(N)}}{\partial\Omega^\alpha} + \mu\left(t\right)\Omega^\alpha
\]
for undetermined functions $\kappa$ and $\mu$. The restricted Hamiltonian then has the form
\[
\hat{h}=\sum_{\alpha}\left[\kappa\left(t\right)\frac{\partial\mathcal{H}_{\textrm{SU}(N)}}{\partial\Omega^\alpha} + \mu\left(t\right)\Omega^\alpha \right]\hat{T}^{\alpha}.
\]
By expanding the definition of $\frac{\partial\mathcal{H}_{\textrm{SU}(N)}}{\partial\Omega^\alpha}$ and comparing with Eq. (\ref{eq:generic_SU(N)_hamiltonian}), we find,
\[
\frac{\partial\mathcal{H}_{\textrm{SU}(N)}}{\partial\Omega^\alpha}=\frac{\partial\left\langle \Omega\right|\hat{\mathcal{H}}\left|\Omega\right\rangle}{\partial\left\langle \Omega\right|\hat{T}^{\alpha}\left|\Omega\right\rangle }=b_\alpha.
\]
If we impose the restriction that $\hat{h}$ produce
the same dynamics as $\hat{\mathcal{H}}$ when the latter is itself 
a linear combination of only the spin operators, it follows that $\kappa\left(t\right)=1$
and $\mu\left(t\right)=0$. The restricted Hamiltonian is therefore 
\[
\hat{h}= \sum_{\alpha=1}^3\frac{\partial\mathcal{H}_{\textrm{SU}(N)}}{\partial\Omega^\alpha}\hat{T}^\alpha.
\]

The classical equations of motion for our restricted dynamics may now
be calculated. Recall once again that we need only consider the first
three $\hat{T}^{\alpha}$ and may therefore 
write the relevant structure constants $f_{\alpha\beta\gamma}$ as
$\epsilon_{\alpha\beta\gamma}$
\[
\begin{alignedat}{1}i\hbar\frac{d}{dt}\hat{T}^{\alpha} & =\left[\hat{T}^{\alpha},\hat{h}\right]\\
 & =\sum_{\beta=1}^3\frac{\partial\mathcal{H}_{\textrm{SU}(N)}}{\partial\Omega^\beta}\left[\hat{T}^{\alpha},\hat{T}^{\beta}\right]\\
 & = i\hbar \sum_{\beta,\gamma = 1}^3 \epsilon_{\alpha\beta\gamma}\frac{\partial\mathcal{H}_{\textrm{SU}(N)}}{\partial\Omega^\beta}\hat{T}^{\gamma}
\end{alignedat}
\]
Taking expectation values of both sides we find,
\begin{equation}
\frac{d\Omega^{\alpha}}{dt}= \sum_{\beta,\gamma = 1}^3 \epsilon_{\alpha\beta\gamma}\frac{\partial\mathcal{H}_{{\rm SU}\left(N\right)}}{\partial\Omega^{\beta}}\Omega^{\gamma},
\label{eq:motion}
\end{equation}
which is precisely Eq. (\ref{eq:restricted_dynamics_final}) for a single site Hamiltonian. Note that Eq~\eqref{eq:motion} can be expressed in terms of the spin Poisson bracket \cite{lakshmanan2011fascinating},
\begin{equation}
\{\mathcal{A}, \mathcal{B}\}=\sum_{\alpha, \beta, \gamma=1}^3 \epsilon_{\alpha \beta \gamma} \frac{\partial \mathcal{A}}{\partial \Omega^\alpha} \frac{\partial \mathcal{B}}{\partial \Omega^\beta} \Omega^\gamma,
\end{equation}
where $\mathcal{A}$ and $\mathcal{B}$ are two functions of ${\bf \Omega}$:
\[
\frac{d{\bf \Omega}}{dt}= \{{\bf \Omega}, \mathcal{H}_{{\rm SU}\left(N\right)}\}.
\]
Moreover, the dynamics may be rewritten directly in terms of angles, rather than expectation values, which permits formulation in terms of canonical Poisson brackets, as outlined in Appendix A of \cite{Hao21}.

\bibliographystyle{apsrev4-2}
\bibliography{ref}